\documentclass[a4paper,11pt]{article}

\usepackage{graphicx,color}
\usepackage{jheppub} 
\usepackage{enumitem}
\usepackage{braket}
\usepackage{slashed}
\usepackage[utf8]{inputenc}
\usepackage{hyperref}
\usepackage{float}
\usepackage{caption}
\usepackage{subcaption}
\usepackage{makecell}
\usepackage{epsfig}
\usepackage{amsmath,amssymb}
\usepackage[normalem]{ulem}
\usepackage{tikz}
\usetikzlibrary{shapes.geometric, arrows}

\hypersetup{
	colorlinks=true,
	citecolor=black,
	filecolor=black,
	linkcolor=blue,
	urlcolor=blue
}

\newcommand{\lr}[1]{\left(#1\right)}

\title{String Thermodynamics In and Out of Equilibrium: Boltzmann Equations and Random Walks}

\author[a]{Andrew R. Frey,}
\author[b]{Ratul Mahanta,}
\author[c]{Anshuman Maharana,}
\author[d]{Francesco Muia,}
\author[d]{Fernando~Quevedo,}
\author[d]{Gonzalo Villa}

\affiliation[a]{Department of Physics and Winnipeg Institute for Theoretical Physics,
University of Winnipeg, 515 Portage Avenue, Winnipeg, Manitoba R3B 2E9, Canada}
\affiliation[b]{INFN, Sezione di Bologna, viale Berti Pichat 6/2, 40127 Bologna, Italy}
\affiliation[c]{Harish Chandra Research Institute, A CI of Homi Bhabha National Institute,
Chattnag Road, Jhunsi, Prayagraj (Allahabad) - 211019, India}
\affiliation[d]{\footnotesize DAMTP,  Centre for Mathematical Sciences,  University of Cambridge, Wilberforce Road,  Cambridge, CB3 0WA, UK}

\emailAdd{a.frey@uwinnipeg.ca}
\emailAdd{mahanta@bo.infn.it}
\emailAdd{anshumanmaharana@hri.res.in}
\emailAdd{fm538@cam.ac.uk}
\emailAdd{fq201@damtp.cam.ac.uk}
\emailAdd{gv297@cam.ac.uk}

\abstract{ We revisit the study of string theory close to the Hagedorn temperature with the aim towards cosmological applications. We consider interactions of  open and closed strings in a gas of D$p-$branes, and/or one isolated D$p$-brane, in an arbitrary number $d$ of flat non-compact dimensions and general compact dimensions. Leading order string perturbation theory is used to
obtain the basic interaction rates in a flat background, which are shown to be consistent with the random walk picture
of highly excited strings that should apply in more general backgrounds. Using the random walk interpretation we infer the structure of more general semi-inclusive string scattering rates and then write down the corresponding Boltzmann equations describing  ensembles of highly excited closed and open strings. We organise the interaction terms in Boltzmann equations 
so that  detailed balance becomes manifest. 
We  obtain  the equilibrium solutions and show that they reduce to previously computed solutions for $d=0$.
We further study the  behaviour of non-equilibrium fluctuations and find explicit analytic expressions for the equilibration rates (and for the number of open strings in $d=0$). Potential implications for an early universe with strings at
high temperatures are outlined. 

}

\begin{document} 
\maketitle
\flushbottom

\section{Introduction}\label{sec:intro}

The high energy regime of string theory remains a mysterious place.
The direct approach to study this limit is to examine the scattering or 
decay of highly excited strings (see, for example,~\cite{Gross:1987kza, Gross:1987ar, Amati:1987wq, Amati:1987uf, Amati:1990xe, Mitchell:1988qe, Gross:1989ge, Amati:1999fv, Manes:2001cs, Manes:2003mw, Manes:2004nd, Veneziano:2004er, Chen:2005ra, Giddings:2007bw, Giddings:2011xs, Hindmarsh:2010if, Skliros:2011si, Skliros:2013pka} or more recently~\cite{Gross:2021gsj, Rosenhaus:2021xhm, Firrotta:2022cku, Bedroya:2022twb, Firrotta:2023wem, DiVecchia:2023frv}).
Alternatively, one can  seek to understand the high-temperature thermodynamics of strings~\cite{Sundborg:1984uk, Tye:1985jv, Alvarez:1985fw, Bowick:1985az, Salomonson:1985eq, McClain:1986id, Sathiapalan:1986db, OBrien:1987kzw, Mitchell:1987hr, Mitchell:1987th, Axenides:1987vi, Atick:1988si, Brandenberger:1988aj, Deo:1988jj, Deo:1989bv, Bowick:1989us, Deo:1991mp, Lowe:1994nm, Lee:1997iz, Abel:1999rq, Barbon:2004dd, Mertens:2015ola, Brustein:2021ifl, Brustein:2022uft, Brustein:2022wiq}, which is closely  related to the interactions
of highly excited string states. This is a much studied subject, and implications for early universe cosmology can be very important~\cite{ Nayeri:2005ck, Brandenberger:2006vv, Brandenberger:2006xi, Frey:2005jk, Frey:2021jyo}.
Yet, many open problems remain. In this article we will address some of these
(touching upon both equilibrium and non-equilibrium aspects) primarily focusing on the Boltzmann
equation approach. We begin by providing a brief summary of our results; for convenience, we use units in which the string length is one throughout the manuscript.

\subsection*{Summary of Results}
\begin{itemize}
\item  Using the standard formulation of kinetic theory and detailed balance we analyse the 
form of equilibrium number densities for strings. The standard result is 
Bose-Einstein and Fermi-Dirac distributions multiplied by the level degeneracies. 
For strings, there is the additional caveat that highly energetic strings (which are very long) can
occupy the entire volume, the  structure of the phase space changes.
This sets the background for obtaining the ``detailed balance" equations for highly excited strings. This is done in section \ref{sec:general-argument}.

\item Making use of results available in the literature~\cite{Amati:1999fv,Manes:2001cs,Chen:2005ra} and  carrying out generalisations, we analyse interactions and decay rates of highly excited strings so as to gather the inputs needed to
set up Boltzmann equations. The interactions/decays are length preserving and  consistent
with the random walk picture of highly excited strings. This is done is section
\ref{sec:decayrates}.
\item We set up Boltzmann equations (along the line of \cite{Lowe:1994nm,Lee:1997iz}, and ~\cite{Copeland:1998na} where a  similar study of a distribution of closed loops was carried out in the context of cosmic strings) in the case that open/closed strings
are in effectively non-compact directions.\footnote{A dimension will be considered effectively non-compact if
it is of large enough size so that typical strings of the ensemble perceive then to be non-compact. Having non-compact directions is important for 
cosmological applications.}
The random walk picture is used as a guideline.
Equilibrium solutions are obtained using 
detailed balance. An interesting feature is that detailed balance is organised length by length
of the strings. This is done is section \ref{sec:boltzmann}.
\item Non-equilibrium dynamics (for  closed strings  and  closed/open admixtures) is analysed by considering perturbations about the equilibrium configurations. Explicit analytic solutions are obtained, the dependence of the damping rates on length of
strings is discussed. There is very little study of  non-equilibrium dynamics of highly excited strings, our study opens up various avenues
for exploration. This is done in section \ref{sec:perturbations}.

\end{itemize}

\subsection*{Review}

  Let us review some  aspects of string thermodynamics. Here, we will focus on basic features of the density of states (for free strings). In some sense, this is complementary
to the Boltzmann equation approach --- we give only a brief discussion so that the reader can connect to the results obtained in the main text.
Also, we will explain here the notion of ``effectively non-compact directions" in string thermodynamics. We refer the reader to
\cite{Mertens:2015ola} for a more detailed review 
and a comprehensive list of references.

It is a simple consequence of the exponential degeneracy of the density of states of a system of strings, $\Omega(E)\sim E^\gamma e^{\beta_H E}$
--- where $E$ is the state energy and $\beta_H\sim \sqrt{\alpha'}$ is the inverse Hagedorn temperature ---
that something interesting occurs as the temperature $1/\beta$ of a system reaches the value $1/\beta_H$.
Indeed, in the formal regime $\beta < \beta_H$ (\textit{i.e}, at temperatures higher than the Hagedorn temperature),
the canonical partition function, $Z\sim \int_0^\infty{dE\, E^\gamma e^{-(\beta-\beta_H) E}}$, diverges exponentially.
Such Hagedorn behaviour was studied in the early days in the bootstrap model for hadrons, e.g.~\cite{Hagedorn:1965st, Hagedorn:1967tlw, Hagedorn:1967dia, Hagedorn:1971mc, Frautschi:1971ij, PhysRevD.5.3231, Cabibbo:1975ig}, and it was shown that for $\gamma<-1$ the microcanonical and canonical ensembles do not agree, due to large energy fluctuations of the latter. 

The issue in string theory is even more intricate.
As was shown in~\cite{Deo:1988jj}, one must be very careful in taking the thermodynamic limit where the energy $E\rightarrow \infty,$ the volume of the system $V\rightarrow \infty$, and $\rho=E/V=\text{constant}$, not only due to the Jeans instability (present in any gravitational system) but also due to the fact that even a single highly excited string can fill 
the entire volume of the space a large number of times.
In  fact, it is compactness that saves the high energy regime of string theory from inconsistencies like a negative specific heat. 
It was argued in~\cite{Brandenberger:1988aj} that \textit{any} background where string propagation can be 
described by a nine dimensional unitary superconformal CFT (supplemented with a time direction)  has single-string density of states $\hat{d}$ in the high energy regime of the form:
\begin{equation}\label{eq:compact-dos}
    \hat{d}(\varepsilon)=\frac{e^{\beta_H \varepsilon}}{\varepsilon}\, ,
\end{equation}
for a individual string of energy $\varepsilon$, and corrections to this formula yield a positive specific heat, and thus a proper thermodynamic behaviour.\footnote{
This quantity is not to be confused with the full density of states of the system, $\Omega(E)$, which determines whether the microcanonical and canonical ensembles agree.} The case where some of the dimensions are (effectively) noncompact is more intricate. 
In this case, the density of states takes the form
\begin{equation}\label{eq:noncompact-dos}
    \hat{d}(\varepsilon)=V\frac{e^{\beta_H \varepsilon}}{\varepsilon^{1+d/2}}\, ,
\end{equation}
where $d$ is the number of effectively non-compact directions and $V$ is their volume in string units (the energies 
are also in string units). 
Using this naively in the high energy regime, one obtains a negative specific heat.\footnote{
As is well known, this density of states gives rise to a system composed of one very large string carrying most of the energy density, surrounded by a bath of very short strings.}

A consistent picture is obtained by requiring that for thermodynamical questions we always put strings
on compact spaces, but depending on the energy scales involved, directions can be effectively non-compact. The difference between \eqref{eq:compact-dos} and \eqref{eq:noncompact-dos} has a simple
explanation in the random walk picture of highly excited strings. A long string of length $l$ 
(which has energy $l$) is a random walk in $d$ dimensions which effectively occupies a volume
$l^{d/2}$. The regime in which the string perceives the space as compact corresponds to
  $l^{d/2} \gg V$. 
Furthermore, because it effectively occupies the entire volume of the space there is no translational zero-mode: the degrees of freedom
are those of internal excitations of the string.  On the other hand, for  $l^{d/2} < V$,
the space is perceived as non-compact. The density of states in \eqref{eq:noncompact-dos}
has a multiplicative relative factor of  $\frac{V}{ l^{d/2}}$ in comparison with \eqref{eq:compact-dos},
corresponding  to degrees of freedom associated with the translational zero mode
of the string \cite{Polchinski:1994mb,Abel:1999rq}.

Similarly, the single open string density of states $d_o(\varepsilon)$ in the torus setup satisfies, in a configuration with $N$ parallel D$p$-branes separated by a distance $X\ll \sqrt{\varepsilon}$~\cite{Abel:1999rq}
\begin{equation}
    d_o(\varepsilon)\sim \begin{cases}
        N^2\frac{V_{\|}}{\varepsilon^{(d-p)/2}}e^{\beta_H \varepsilon}\, & R\gg \sqrt{\varepsilon}\, ,\\    
        & \\
        N^2\frac{V_{\|}}{V_{\perp}}e^{\beta_H \varepsilon}\, & R\ll \sqrt{\varepsilon}\, .
    \end{cases}
\end{equation}
where $N$ is the number of D-branes, $V_\|$ $(V_\perp)$ is the volume parallel (transverse) to the branes, and $d$ is the number of effectively noncompact dimensions (with $d-p$ orthogonal to the branes) with typical radius $R$. 

We thus conclude that the notions of density of states and effectively noncompact dimensions depend on the average length of the string. 
Thus, the thermodynamics will be described differently in the different energy regimes. In this paper we will describe how all of these densities of states appear as solutions of the Boltzmann equations, with the appropriate interaction rates for highly excited strings.
We will not, however, aim to describe the transition between regimes, since it is during these transitions that the canonical ensemble fails~\cite{Barbon:2004dd}.

\subsection*{Key Ingredients}

   Now, we describe the key ingredients of our analysis.

\begin{itemize}
    \item The Worldsheet 
    
 Interactions are crucial for writing Boltzmann equations. 
The primary approach to calculating scattering (or decay) amplitudes for high energy string states is string perturbation theory on the worldsheet. For example,~\cite{Amati:1999fv} computed the emission of massless states for highly excited strings, and, in a similar fashion, \cite{Manes:2001cs} calculated decay rates of long strings by emission of strings of arbitrary mass (generalized in~\cite{Chen:2005ra} to include supersymmetric strings and compactification effects). 
Further examples include~\cite{Manes:2003mw,Manes:2004nd} which worked out the cross section for low-energy tachyons scattering off a highly excited string. In this paper, we will review the worldsheet calculation of decay rates from \cite{Manes:2001cs} and clarify the work of \cite{Chen:2005ra} including KK and winding number.

The worldsheet has also been used to study the  single string density of states in the microcanonical ensemble from explicit state counting~\cite{Mitchell:1987th} or by studying the singularity structure of the free energy~\cite{Deo:1988jj,Deo:1989bv,Deo:1991mp}, agreeing in the conclusion of Eqs.~\eqref{eq:compact-dos} and~\eqref{eq:noncompact-dos}. As pointed out by those authors, there is a risk of inconsistency in this approach: the calculations are performed for a free string, while ergodicity (fundamental for the microcanonical approach) requires an interaction among the string constituents. In this paper, we give a short but general argument for why their results apply to the general case of interactions in string perturbation theory, as long as the modifications to the  string spectrum vanish in the limit of $g_s \to 0$ (\textit{i.e} as long as string perturbation theory holds). 

    \item Random walks

    It has long been argued that highly excited strings form random walks (see e.g.~\cite{Mitchell:1987th}).
    A simple way to see this is through entropic arguments: one can show that the leading order contributions to the entropy of a highly excited state and a random walk coincide.
The worldsheet scattering amplitudes of~\cite{Manes:2003mw,Manes:2004nd} mentioned above yield a form factor consistent with a random walk structure for long strings.
    
    Similar conclusions to those in the worldsheet approach can be reached using random walk arguments (see e.g.~\cite{Barbon:2004dd}) to compute the single-string density of states.
    In this paper, we explain the random walk interpretation of the decay rates of long, fundamental strings.
    This random walk interpretation serves as a guide for the form of higer-point functions and hybrid interactions in situations where a  worldsheet derivation is not available. We check the validity of these by showing that they satisfy detailed balance for the known equilibrium distribution. 

    \item The Boltzmann equation

   Armed with the form of the interaction rates, we will write down Boltzmann equations following the
approach of~\cite{Lowe:1994nm,Lee:1997iz} and ~\cite{Copeland:1998na} in the context of cosmic strings.
Note that this approach yields Eq.~\eqref{eq:compact-dos} for the density of states at $d=0$, hence  confirming validity of the result in the presence of interactions.
We will obtain Boltzmann equations (for both open and closed strings)  when some of the directions are effectively non-compact (corresponding to the system at lower energies).
These equations allow us to probe out-of equilibrium phenomena, of which very little has been explored in the case of strings.

\end{itemize}

\subsection*{Organisation of the Paper}

This paper is organized as follows: in Section~\ref{sec:general-argument}, we settle our notation with a short introduction to the generalization of standard kinetic theory to string theory.
Through a textbook argument, we introduce the notion of detailed balance, which illustrates (in a background independent way) why the computations for equilibrium configurations performed for free strings agree with those computed in string perturbation theory.
Unsurprisingly, these equilibrium configurations are given by Bose-Einstein/Fermi-Dirac distributions, corrected by an exponential density of states.

In Section~\ref{sec:decayrates}, we summarize the computation of semi-inclusive averaged interaction rates to leading order in string perturbation theory, using the important results of~\cite{Manes:2001cs}.
We find that the behaviour of the typical decay rate is such that strings behave non-relativistically and that winding modes are efficiently distributed proportionally to the mass of the strings.
This justifies using the length (equivalently, mass) of the string to describe the thermodynamics at sufficiently high level.
Furthermore, we provide a random walk interpretation of the interaction rates, which allows us to conjecture the form of higher-order interaction rates.
To the best of our knowledge, many of these interaction rates have not appeared elsewhere in the literature.
These conjectures satisfy the non-trivial test of consistency with detailed balance for the equilibrium distribution found for lower-order interactions in string perturbation theory.

Section~\ref{sec:boltzmann} contains the main result of the paper. Using the results above, we pose the Boltzmann equations describing three different regimes, and find equilibrium distributions.
First, we consider the thermodynamics of closed strings in $d$ effectively noncompact spatial dimensions, finding\footnote{Subscripts $c,o$ respectively correspond to closed and open strings in our notation.}
\begin{equation}\label{eq:onlyclosed}
     n_c(l) = V \frac{\kappa_b}{\kappa_a}\frac{e^{-l/L}}{l^{1+d/2}} \, ,
\end{equation}
where $n_c(l)dl$ is the number of strings with lengths\footnote{Throughout the text, we exchange the length of the strings by their masses, as they are the same in string units.} between $l$ and $l+dl$, and discuss why this is in agreement with previous results upon identification of $1/L=\beta-\beta_H>0$.
The constants $\kappa_{a,b} \sim g_s^2$ are $d-$dependent numbers which do not affect our discussion.

Later, we introduce D-branes, and force the open and closed strings to lie within a $d$-dimensional, effectively noncompact, space-filling brane or, equivalently, densely spaced D$p$-branes ($p<d$).
We find that the equilibrium distributions of strings are
\begin{equation}\label{eq:confined}
     n_o(l)=\frac{2a \kappa_b N^2V_\parallel}{b \kappa_aV_\perp}e^{-l/L}\, , \qquad 
     n_c(l) = V \frac{\kappa_b}{\kappa_a}\frac{e^{-l/L}}{l^{1+d/2}} \, ,
\end{equation}
where $N$ is the number of parallel, effectively overlapping D$d$-branes in the $d$ directions.
These span a worldvolume $V_\parallel$, and the transverse volume is denoted $V_\perp$, which together yield a total volume $V=V_\parallel \cdot V_\perp$, and $a$ and $b$ are $\mathcal{O}(g_s)$ and $d-$dependent computable quantities.

Lastly, we let the strings probe the dimensions transverse to the brane, thus describing the thermodynamics of a D$p$-brane in a $d+1$-dimensional spacetime.
We find that the equilibrium distributions are now given by
\begin{equation}
     n_o(l)=\frac{2a \kappa_bN^2V_\parallel}{b \kappa_a}\frac{e^{-l/L}}{l^{(d-p)/2}} \, , \qquad 
     n_c(l) = V \frac{\kappa_b}{\kappa_a}\frac{e^{-l/L}}{l^{1+d/2}} \, .
\end{equation}

In Section~\ref{sec:perturbations} we study the behaviour of these equilibrium distributions under fluctuations.
For the $d=0$ case in absence of branes, we find an analytic solution for the behaviour of linear fluctuations, which decay with a length-dependent rate $\Gamma(l)=\kappa (l^2/2+l L)/V$.
With knowledge of the explicit solution, we study the origin of this qualitative behaviour of $\Gamma (l)$ and argue for its expression in more complicated setups.
We find length-dependent equilibration rates that are sensitive to the number of non-compact directions, the total energy of the gas and the string coupling.

Lastly, we conclude in Section~\ref{sec:conclusions} with a summary and future directions.
This approach to string thermodynamics is particularly powerful because it can describe the evolution of the thermal system when it falls out of equilibrium, providing us with computational tools to study a putative Hagedorn phase in the early Universe,
which we leave for a future paper~\cite{Frey:2024in}.

\section{Stringy kinetic theory}\label{sec:general-argument}

In this section we summarize the setup for a stringy version of kinetic theory (see e.g.~\cite{Tong:notes} for an introduction to standard kinetic theory) and give an argument for why the density of states gives the string number density even in presence of interactions. These ideas will lead us to the detailed balance equations in the later sections.

Consider a probability density for a single string in a generalized phase space, $f(r,k,N,\sigma)$, which apart from the usual position and momentum variables $r,k$ in the non-compact directions,\footnote{Note there is no position dependence along the compact directions, since the string is occupying the whole space.} includes the oscillator level $N$, and other discrete variables $\sigma$ which in toroidal compactifications include winding and KK modes.
For ease of notation, in this section we will write $f(E_l,l)$, where in the example of closed strings in a toroidal background we have
\begin{equation}\label{eq:energy}
    E_l^2=k^2+\frac{2}{\alpha'}(N_L+N_R-2)+\sum_{i=1}^{d_c}\lr{{\lr{\frac{n_i}{R_i}}^2+\lr{\frac{\omega_iR_i}{\alpha'}}^2}}\, ,
\end{equation}
where $k$ is the spatial momentum in the (effectively) noncompact directions, $N_L$ and $N_R$ are left and right moving oscillator levels, respectively, and ${n_i,\omega_i}$ denote the KK and winding modes along the $i^{th}$ direction, of radius $R_i$.
Throughout the text, we will refer to the length of the string as its proper length, defined by $M=\mu l$, where $\mu$ is the tension of the string and $M$ its mass, obtained from~\eqref{eq:energy} through $E^2=k^2+M^2$. 
In the rest of the text, we will use the chemistry notation $l\rightleftharpoons x+y$ to indicate the decay of a string of length $l$ to a pair of strings of lengths $x$ and $y$, or the absorption of a string of length $x$ by a string of length $y$ to yield a string of length $l$. 
The amplitude of any such process will be given by 
\begin{equation}\label{eq: }
    |{\langle \Phi_{x} | V_y(k) | \Phi_{l}}\rangle |^2\equiv \mathcal{A} \, ,
\end{equation}
where we have considered particular states (\textit{i.e.} with fixed $r,k,N,\sigma$), and have made use of the state-operator correspondence to write the amplitude in this way, because it is this form that we will use in Sec.~\ref{sec:decayrates} to compute the interactions.

This reaction will contribute the following term in the Boltzmann equation for $f(E_l,l)$:
\begin{equation}
    \frac{\partial f(E_l,l)}{\partial t} \supset \sum_{\sigma_x, \sigma_y}{\int_{\mathbb{R}^p}{d^pk_x \, d^p k_y\, \mathcal{A} \lr{ f_xf_y\lr{1\pm f_l}-f_l\lr{1\pm f_x}\lr{1\pm f_y}}}\delta(...)} \,,
\end{equation}
where $\delta(...)$ takes into account several conservation equations, like energy or, in the case of only closed strings, winding number.
The equilibrium configuration $\partial f_{eq}/\partial t=0$ can be found by the requirement of \textit{detailed balance}, i.e that the terms of all the individual reactions that contribute to the Boltzmann equation cancel out separately.
It is easy to see that this yields a Bose-Einstein/Fermi-Dirac distribution for the individual strings:
\begin{equation}
    f_{eq}(r,k,N,\sigma)=\frac{1}{e^{\beta E}\mp 1}\, ,
\end{equation}
where $\beta$ is not fixed yet.
This result is not useful for several reasons. 
First of all, it is simply the canonical ensemble statement that all microstates with the same energy have the same probability.
It is, however, very unlikely that in a thermodynamic setup we know about the particular state of a system, or (as lower dimensional observers) its windings.
The best we can aim at is to know the mean length (equivalently, mass) of the individual strings. 
Nicely, for fixed $k$, all such strings have the same energy, and we can thus write
\begin{equation}\label{eq:eq-dist}
    f_{eq}(r,k,l)=\frac{\mathcal{N}(l)}{e^{\beta E}\mp 1}\, .
\end{equation}

This is how the density of states at a given mass $\mathcal{N}(l)$ shows up in our considerations.
In what follows, we will set up Boltzmann equations for the \textit{typical} strings (\textit{i.e: averaged within an oscillator level}), for which we will argue that the energy is dominated by the level (as opposed to the combination of level plus winding and KK modes), and thus can be described by a net number\footnote{Note that we depart here from the usual statistical mechanics notions, where the density of particles is used. We use the net number for consistency with previous work~\cite{Lowe:1994nm,Lee:1997iz} and because for $d=0$, where the strings occupy the whole volume, the definition of density becomes subtle.} phase space density $n(l)dl$.
We will use the high energy limit of Eq.~\eqref{eq:eq-dist} (analogously, Eq.~\eqref{eq:compact-dos} or~\eqref{eq:noncompact-dos}, as appropriate) as a consistency check for our results.
\\

Eq.~\eqref{eq:eq-dist} is a simple consequence of ordinary statistical mechanics, reproducing the Bose-Einstein/Fermi-Dirac distribution to the quantity of interest in string thermodynamics, $n(l)$. 
It is worth noting that this expression applies for strings at any energy, and no divergences (except for the well known Bose-Einstein condensation phenomenon for massless strings) appear in the equation.
This is different than the expression one would obtain from Eq.~\eqref{eq:compact-dos} because that one only applies when the energy is dominated by the mass (equivalently, length) of the string.
In the appropriate high energy regime, of course, the results are equivalent.\footnote{The results differ in $d-$dependent constants because $f(r,k,l)$ and $n(l)$ are not exactly the same quantity, and differ by $e.g.$ integration over all external momenta at given mass.}

More interestingly, this equilibrium distribution together with the exponential form $e^{\beta_H l}$ of the density of states predicts a Bose-Einstein/Fermi-Dirac distribution for massless fields at an effective temperature $1/\beta<1/\beta_H$. 
We can point out two implications of this:
\begin{itemize}
    \item As long as our description is valid, the temperature of the massless fields never reaches the Hagedorn temperature.
    This is because once the strings reach energies of order one, oscillator modes govern the thermodynamics. 
    \item Due to the $N^2$ degeneracy and the fact that $n_o(l)\sim ln_c(l)$, open strings dominate the ensemble. 
    These observations suggest interesting features of a reheating period involving the Hagedorn phase.
\end{itemize}

The argument exposed in this section is purely kinematic. 
It is easy to see that, upon summing over all final configurations and averaging over all possible initial conditions, the statement of detailed balance applies in the same way for higher point functions, with more terms.
We have thus shown that the distribution in Eq.~\eqref{eq:eq-dist} holds at all orders in string perturbation theory. 
Note also that the argument is background independent, although of course the difficulty now lies in computing $\mathcal{N}(l)$, which is known for toroidal compactifications but not necessarily in more complicated backgrounds.\footnote{For technical reasons, we will  restrict our analysis to toroidal backgrounds for various explicit computations in the paper.
It would be interesting to generalize this approach to more realistic compactifications.} 

  Before closing this section, we note that the ideas in this section will help us to
set up the detailed balance conditions that we will use to obtain equilibrium solutions
to the Boltzmann equation (in section \ref{sec:boltzmann}).

\section{General results in decay rates}\label{sec:decayrates}

In the previous section, we have argued how the requirement of detailed balance in the Boltzmann equations gives rise to the well known Bose-Einstein/Fermi-Dirac distributions. 
There is a natural question: under what conditions can the strings reach equilibrium? 
In ordinary thermodynamics, one simply waits long enough until equilibrium is reached, provided there is some interaction that allows the system to move through phase space.
In situations with gravity, this is no longer true, and the dynamics of spacetime do not allow us to ``wait long enough", unless the interaction rates are much faster than the gravitational effects on the system.
Examples include decoupling of species due to the expansion of the Universe or overdensities due to gravitational collapse.
Knowledge of interaction and equilibration rates is thus essential to assess whether equilibrium thermodynamics appropriately describes a gravitating system.
\\

A worldsheet computation of an amplitude involving highly excited strings is in general complicated, due to the degeneracy and the complicated spin structure of the state (although progress has been made in this direction~\cite{Hindmarsh:2010if, Skliros:2011si, Skliros:2013pka}).
However, a very simple observation allows for the computations that are relevant for us: in a thermal ensemble, the interaction rate is dominated by the \textit{typical} string.
Phrased in a different way, since all oscillator states have the same probability, in a thermodynamic system we cannot be sure of which is the particular initial state of an interaction, and thus we must average over all such initial states.
This observation was first used in~\cite{Amati:1999fv} to compute the emission of massless strings from very massive ones.

In this section we study interaction rates of the typical string from a worldsheet point of view, closely following~\cite{Manes:2001cs}.
We also consider compactification effects~\cite{Chen:2005ra}, and include further insights on absorption rates.
The main point is that the form of the decay rates at sufficiently high level predicts a distribution of strings that is completely determined by their length, $n(l)$.
Furthermore, we give a random-walk interpretation for the decay rates. 
This allows us to conjecture the form of the hybrid interactions, mixing open and closed strings, and having this we will be able to pose a system of Boltzmann equations for $n(l)$.

\subsection{Averaged 3-point amplitude}

The idea is as follows. 
We want to study the process of decay of a highly excited string at fixed level $N$, to a string in a specific state determined by a vertex operator $V_1(k)$, and another massive string at fixed level $N_2<N$.
In the case of a compactification, we also fix winding and KK momentum of the strings.
Because we are considering a thermal ensemble, we average over all possible initial configurations at this level.
Summing over all possible final states at fixed level $N_2$, the amplitude for the process is given by
\begin{equation}\label{eq: smatrix}
    \frac{F(N,N_2)}{\mathcal{G}(N)}\equiv \frac{1}{\mathcal{G}(N)}\sum_{\Phi_N} \sum_{\Phi_{N_2}}|{\langle \Phi_{N_2} | V_1(k) | \Phi_{N}}\rangle |^2 \, , 
\end{equation}
where $\mathcal{G}(N)$ is the oscillator degeneracy of the level $N$ (not to be confused with the number of states at a given mass $\mathcal{N}(l)dl$), and the different $|\Phi_N\rangle$ account for the different states at a level $N$. 
Through the insertion of projectors at levels $N$ and $N_2$, this expression can be written as a trace over oscillator modes\footnote{Importantly, we are not summing over windings/KK modes, and because of this reason the calculation is the same in the compact and noncompact case.}
\begin{equation}\label{eq: trace}
    F(N,N_2) = \oint_C \frac{d\, z}{z}z^{-N} \oint_{C_2} \frac{d\, z_2}{z_2}z_2^{-N_2} \text{Tr} \left[z^{\hat{N}}V_1^\dag (k,1) z_2^{\hat N}V_1(k,1)\right] \, ,
\end{equation}
where the projectors are contour integrals around zero, and $\hat{N}=\sum_{n=1}^\infty{\alpha_{-n}\cdot \alpha_n}$ is the number operator. 
In passing, we note that an appropriate redefinition renders this formula and the upcoming discussion valid for absorption of strings (in which case $N_2>N$). 
The S-matrix element in Eq.~\eqref{eq: smatrix} is the same upon exchange $V_1 \rightarrow V_1^\dag$. 
This implies, using the cyclic property of the trace in Eq.~\eqref{eq: trace}, that for processes involving an initial state at level $N$ and a final state at level $N_2$, 
\begin{equation}
    F_{em}(N,N_2)=F_{abs}(N_2,N)\, ,
\end{equation}
which illustrates the crossing symmetry of the amplitude.
\\

Using coherent state techniques \cite{Green:1987sp}, it can be argued that, in $D$ spacetime dimensions,
\begin{gather}
\begin{split}
    F(N,N_{2})=\oint \frac{d \omega}{\omega}\omega^{-N}& f(\omega)^{2-D} \mathcal{I}_{N-N_2}(\omega) \,, \\ 
    \mathcal{I}_n \equiv \oint \frac{dv}{v} \, v^n\langle &V'^\dag(k,1) V'(k,v) \rangle  \, ,
\end{split}
\end{gather}
where $f(\omega)$ is related to the partition function (see~\cite{Manes:2001cs}) and the prime on the vertex operator indicates a subtraction of a zero-mode that is not relevant for our discussion. 
The key observation of~\cite{Manes:2001cs} is that for known level $N_1$ of the product string of known state, the functions $\mathcal{I}_n (\omega)$ satisfy recursion relations which depend only on $N_1$. 
This leads to the final result 
\begin{gather}\label{eq:manes-amplitude}
\begin{split}
    F(N,N_2)=-\sum_{p=1}^A \frac{\partial \mathfrak{n}}{\partial p}\mathcal{G}\left[\mathfrak{n}\right] & +F_{NU}(N,N_2) \, , \\
    \mathfrak{n}\equiv N-(N-N_2)p+& (N_1-1)(p^2-p)\, , \\
    F_{NU}(N,N_2)\equiv \oint \frac{d \omega}{\omega}\omega^{-N}f(\omega)^{2-D} & \mathcal{I}_\nu (\omega) \omega^{\nu A +(N_1-1)(A^2+A)} \, ,
\end{split}
\end{gather}
where $\nu \equiv (N-N_2) \, \text{mod} \, 2(N_1-1)$, and $A=1$ for normalized states and otherwise is the leading coefficient of the OPE of the vertex operators.
As checked in~\cite{Manes:2001cs}, except for $N_1=1$, it is the universal contribution that dominates\footnote{Technically, the non-universal part $F_{NU}$ can become important for $M_1\geq 2M_2$ for soft decays $M\simeq M_1+M_2$. For our purposes, if this were the case, we could trade $M_1$ by $M_2$ (\textit{i.e}: assume we know the state of $N_2$ instead of $N_1$) and carry the calculations through.} due to the degeneracy of the levels.
We will only consider this part in what follows, except for massless strings, where an exact solution was computed in~\cite{Amati:1999fv}.

\subsection{Interaction rates}

The amplitude shows crossing symmetry, but the notion of interaction rate does not.
In this subsection we will be careful in understanding what is the dominant product of the decay of a typical string and of the inverse process, the fusion of two typical strings.
This will be the dominant type of string in a thermal ensemble.
We will see that typical decay products are strings with small noncompact momentum in the center of mass frame, and that KK and winding modes are distributed proportionally to the length of the string, in such a way that the energy is efficiently converted into oscillator modes.
In turn, this implies that the mass of the string is essentially given by the oscillator level, up to small corrections, and thus that the energy of highly excited strings is determined by the level. 

\subsubsection*{Emission}

Knowing the amplitude, we can write down the total averaged semi-inclusive decay rate for the process $N\rightarrow V_1+N_2$, where $N$ ($N_2$) is a typical string at level $N$ ($N_2$), and the specific state of $V_1$, at level $N_1$, is known.
Taking into account relativistic normalization of the states and integrating over final external momenta after considering momentum and energy conservation, this is given by
\begin{gather}\label{eq:particular-rate}
\begin{split}
    \Gamma_o\lr{N \rightarrow V_1+N_2}=& A_{d+1} \frac{g_s}{M^2}\frac{F(N,N_2)}{\mathcal{G}(N)}k^{d-2} \, , \\
    \Gamma_{c}\lr{N \rightarrow V_1+N_2}=A_{d+1} \frac{g_s^2}{M^2} &\frac{F(N_L,N_{2,L})}{\mathcal{G}(N_L)}\frac{F(N_R,N_{2,R})}{\mathcal{G}(N_R)}k^{d-2} \, ,
\end{split}
\end{gather}
where $M$ ($M_i$) is the mass of the initial (outgoing) string, $k$ is the outgoing string
momentum,\footnote{Importantly, this quantity is fixed by conservation of energy. We use it as a book-keeping device to study the deviation from length conservation.} we have split the closed string amplitude into left and right movers, and 
\begin{equation}
    A_{d+1}=\frac{2^{-(d+1)}\pi^{1-d/2}}{\Gamma\lr{\frac{d}{2}}}\, .
\end{equation}

These formulae are however not the quantities we are interested in, since they assume knowledge of the particular state of one of the products.
However, because we are considering only the state-independent part of the amplitude (recall Eq.~\eqref{eq:manes-amplitude}), we can include all the product strings at level $N_1$ simply multiplying by the degeneracy $\mathcal{G}(N_1)$.
If, on top of this, both of the product strings are sufficiently long, the dominant contribution to Eq.~\eqref{eq:manes-amplitude} is given by the first term.
Within these approximations (which we use for an analytic understanding but can be easily lifted up if one desires to obtain further accuracy, or is interested in smaller excitation numbers), we get an expression for the decay rate of strings at levels $N_1,N_2$ from a typical string at level $N$:
\begin{equation}\label{eq:rate-N}
\begin{split}
    \Gamma_o\lr{N \rightarrow N_1+N_2}\simeq A_{d+1}\frac{g_s}{M^2}\Tilde{N}\Tilde{\mathcal{G}}k^{d-2} \, , \\
    \Gamma_{c}\lr{N \rightarrow N_1+N_2}\simeq A_{d+1} \frac{g_s^2}{M^2}\Tilde{N}_L\Tilde{\mathcal{G}}_L\Tilde{N}_R\Tilde{\mathcal{G}}_Rk^{d-2} \, , \\
    \Tilde{N}\equiv (N+1-N_1-N_2)\, , \qquad \Tilde{\mathcal{G}}\equiv  \frac{\mathcal{G}(N_1)\mathcal{G}(N_2)}{\mathcal{G}(N)}\, .
\end{split}
\end{equation}
In principle, this is all we need to formulate Boltzmann equations for probability densities $f(k,N,\sigma)$.
However, further study of these expressions allow us to get an idea of what the typical decay strings look like in the ensemble and, as we will see, will allow us to extract simple, analytic expressions describing the decay of the typical string which only depend on its length and the number of noncompact dimensions (even in the presence of winding! See Appendix~\ref{sec:winding-modes}).

In the main text, we quickly review the arguments of~\cite{Manes:2001cs} to find that in noncompact dimensions the typical decay approximately conserves length, from which we infer that the typical string is nonrelativistic in a frame in which the gas has zero net momentum. 
In Appendix~\ref{sec:winding-modes}, we carefully consider the compact case with windings and KK modes, inspired by~\cite{Chen:2005ra}, and show that one can safely ignore the effects of the compact directions at sufficiently large oscillator number.
In passing, we comment on an important mass dependence that was missed in~\cite{Chen:2005ra} and that is the key for a random walk interpretation.
The fact that in Sec.~\ref{sec:boltzmann} it is this corrected contribution that gives the right equilibrium solution, agreeing with Eq.~\eqref{eq:eq-dist} at high energies, supports the claim that our result is correct.
\\

The interplay between the degeneracies and $k$ (which is not a free parameter, but is instead related to the masses by conservation of energy) is what gives the features of the typical interactions.
As argued in~\cite{Manes:2001cs}, in absence of winding, the oscillator degeneracies give rise to a Maxwell-Boltzmann distribution for the \textit{mass defect} $t\equiv M-M_1-M_2$ (\textit{i.e:} the amount of kinetic energy created in the process), which peaks around the mass of the fundamental string. 
This implies that conservation of length is a good approximation for these interactions.
The way to see this is as follows. 
First, recall the Hardy-Ramanujan expression for the degeneracy of the oscillators at sufficiently large ($N\gtrsim 5$) level,
\begin{equation}
    \mathcal{G}(N)\sim (2\pi T_H)^{-(D+1)/2}N^{-(D+1)/4}e^{\sqrt{N/\alpha '}/ T_H}\, .
\end{equation}
Introducing this in $\Tilde{\mathcal{G}}$, it follows from the exponential dependence that $t\sim \mathcal{O}(T_H)\ll \sqrt{N/\alpha '}\simeq M$ and thus that the typical strings are very nonrelativistic, with (as can be easily shown from conservation of energy and momentum) 
\begin{equation}
    t\sim \frac{k^2}{2}\frac{M}{M_1 M_2}\, .
\end{equation}
This allows us to write the expression for $\Gamma$ in terms of $t$ (which in turn determines $N_2$), $N$, and $N_1$.
\begin{gather}\label{eq:notfull-rate}
\begin{split}
    \Gamma_o\lr{N \rightarrow N_1+N_2}\simeq & \frac{g_s}{M_1 M_2}\, t^{d/2-1}e^{-t/T_H} \, , \\
    \Gamma_{c}\lr{N \rightarrow N_1+N_2}\simeq g_s^2 & \lr{\frac{M}{M_1M_2}}^{d/2+1}t^{d/2-1}e^{-2t/T_H} \, .
\end{split}
\end{gather}
This implies that the equipartition principle is satisfied in the noncompact directions,
\begin{equation}
    \langle t\rangle=\frac{d}{2}T_H\, ,
\end{equation}
as can be seen from computing the expectation value of $t$ in this distribution.
\\

We have thus found the probability of decay of a string at level $N$ to a pair of strings at levels $N_1$ and $N_2$.
Still following~\cite{Manes:2001cs}, let us now ask the following questions: what is the total production rate of strings at level $N_1$ from strings at level $N$?
And what is the most likely $N_2$ arising from such a decay?
The answer is obtained by summing Eqs.~\eqref{eq:notfull-rate} over all possible $N_2$ through $dN_2\sim M_2dt$.
The sum is well approximated by an integral, and (up to a numerical constant) this evaluates $t$ at the saddle $t\sim \mathcal{O}(T_H)\ll \text{min}\lbrace M,M_1,M_2\rbrace$, implying $M_2\simeq M-M_1$.
Because departure from this value of $t$ is exponentially supressed, we conclude that decays tend to preserve length.

Lastly, we have computed the rate in terms of the level of the product. 
However, since we will be characterising the strings in terms of their length,\footnote{One can think of the quantites $\Gamma_{o,c}$ in Eq.~\eqref{eq:notfull-rate} as shorthand for $\Gamma/\Delta N_1 \Delta N_2$. We have then summed to get $\Gamma/\Delta N_1=\sum_{N_2}(\Gamma/\Delta N_1 \Delta N_2)\Delta N_2\simeq \int(\Gamma/\Delta N_1 \Delta N_2) M_2 \, dt$, and then expressed $\Gamma/\Delta N_1=(1/M_1)(d\Gamma/dM_1)=(1/M_1)(d\Gamma/dl')$, where the last equality follows in string units.}
we need to convert $dN_1=M_1dM_1$ to find the total number of strings produced with masses between $M_1$ and $M_1+dM_1$:
\begin{gather}\label{eq:decay-rates}
\begin{split}
    \frac{d \Gamma_o}{d l'}\sim &  \, g_s \, , \\ 
    \frac{d\Gamma_{c}}{dl'} \sim  \, & g_s^2 l \lr{\frac{l}{l'(l-l')}}^{d/2} \, .
\end{split}
\end{gather}
where we have defined $l'\equiv M_1/M_s^2$ and $l\equiv M/M_s^2$, work in string units $M_s=1$, and the proportionality constants are independent of $l$ or $l'$.

\subsubsection*{Random walk interpretation}
The results in Eq.~\eqref{eq:decay-rates}, first derived in \cite{Manes:2001cs}, have an appealing semiclassical explanation.
Let us carefully discuss it in this section, since we will later use the intuition developed in the semiclassical picture to argue for the shape of complicated interaction rates.
Importantly, we will discuss the \textit{closed} random walk interpretation of the decay rate of closed strings, which is essential for detailed balance and thus to obtain a correct equilibrium distribution.

In the open string case, Eq.~\eqref{eq:decay-rates} reproduces the usual result that the typical string in a space-filling brane radiates all masses with equal probabilities.
This has the interpretation that the typical string lying along the D-brane can break equally well through any point.
\\

Before discussing the decay rate of a closed string, let us discuss some relevant notions regarding open random walks.
Two points separated by internal length $l$ of an open random walk are separated by an average distance $l^{1/2}$ in a $d$-dimensional space, and the random walk thus fills a volume $l^{d/2}$ in $d$ effectively
noncompact dimensions (i.e, dimensions larger than $l^{1/2}$ in string units). 
Therefore, the probability $P(l)$ that an open random walk of length 
$l$ closes is $P(l)\propto l^{-d/2}$ in $d$ noncompact spatial dimensions
(or $1/V$ if all dimensions are compact), which is the geometric probability that
the second end point is at the same position as the first end point within 
the volume filled by the string.

So what is the probability that a \textit{closed} random walk of length $l$ self-intersects?
Fixing a point of the random walk, we can ask what is the probability of a point at a distance $l'$ to touch this point.
But this is the same question as asking what is the probability that a point at a distance $l-l'$ touches this point.
The decay rate should thus
proportional to the probability that strings of length 
$l'$ and $l-l'$ both close, given that the parent string also closes (which we will denote by $P(l'\& (l-l')|l)$).
We have $P(l'\& (l-l')|l)=P(l'\& (l-l'))/P(l)\sim P(l')P(l-l')/P(l)$ since the
two smaller loops are approximately independent for long strings. 
But this is precisely given by
\begin{equation}\label{eq:prob-closed}
    \lr{\frac{l}{l'(l-l')}}^{d/2}\, ,
\end{equation}
(c.f: Eq.~\eqref{eq:decay-rates}), and the pre-factor $l$ in the decay rate simply shows that the point we have chosen can be any point along the string.
This shows that decay of a semiclassical string occurs upon self-intersection, and that such highly excited strings form closed random walks.
In addition, let us comment on the small string limit. 
It is clear that for $l'/l\ll 1$, the probability of self-intersection of a closed random walk reproduces that of an open random walk. 
This is to be interpreted as the small piece of string not being able to tell whether it belongs to an open or a closed string.
Importantly, this is the behaviour suggested by~\cite{Lowe:1994nm,Lee:1997iz} to generalize the Boltzmann equation approach to $d\neq 0$.
The reason why did this not work is because the closed strings are closed random walks, and the probability of decay in $d$-noncompact directions is Eq.~\eqref{eq:prob-closed} as opposed to $1/l^{d/2}$.
A similar form as Eq.~\eqref{eq:decay-rates} for the interaction rate was argued in~\cite{Copeland:1998na} assuming that the strings form random walks, and imposing detailed balance.
The calculation of~\cite{Manes:2001cs} thus gives a microscopic explanation for their assumptions, even though~\cite{Copeland:1998na} consider field theory strings.
In a similar spirit, it would be interesting to find a worldsheet computation for the interaction rates we pose in Section~\ref{sec:rws} from random walk arguments.
Consistency with detailed balance of these more complicated interaction rates support this picture.
\\

To further illustrate the random walk interpretation of the interaction rates, let us now discuss open strings on D$p$-branes.
Suppose that, instead of moving in $d$ dimensions as in the derivation of Eq.~\eqref{eq:decay-rates}, the open string end points 
are confined in the worldvolume of a D$p$-brane.
Then the phase space factor in 
Eq.~\eqref{eq:particular-rate} is modified to $k^p$.
Tracing appropriately the factors, we find equipartition of kinetic energy in the worldvolume and a decay rate given by
\begin{equation}
    \frac{d \Gamma_o}{dl'}\sim g_s \lr{\frac{l}{l'(l-l')}}^{(d-p)/2}\, ,
\end{equation}
for one open string splitting into two open strings.
This relation has the random walk interpretation that the open string in the ($d-p$) directions transverse to the brane is a closed random walk (since both endpoints must touch the brane).

We will use the random walk interpretation of long string amplitudes 
extensively later in this section.

\subsubsection*{Absorption}

The simple argument we used when studying the derivation of~\cite{Manes:2001cs} yields the following interaction rates for absorption,
\begin{gather}\label{eq:absorption}
\begin{split}
    \Gamma_{o}\lr{N_1+N_2 \rightarrow N}=\frac{2\pi g_s}{8E_1 E_2 E_{M} V}\frac{F(N_1,N)}{\mathcal{G}(N_1)}\delta (E_1+E_2-M) \, , \\
    \Gamma_{c}\lr{N_1 +N_2\rightarrow N}=\frac{2\pi g_s^2}{8E_1 E_2 E_{M}V}\frac{F(N_1,N)^2}{\mathcal{G}(N_1)^2}\delta (E_1+E_2-M) \, .
\end{split}
\end{gather}
It is here where the absorption rate becomes different, since averaging over initial states of mass $M_2$ is trivial, and the phase space integration is different.
In particular, note that there are no phase space terms of the form $k^{d-2}$ because there is only one integral that goes away due to momentum conservation. Thus, terms depending on both the levels and the dimensionality will not appear in this case, and they should not, since we understood them before as random-walk probabilities for self-interaction.
\\

Since $F(N_1,N)\simeq (N+1-N_1-N_2)\mathcal{G}(N_1)\simeq 2\, M_1\, M_2\, \mathcal{G}(N_1)$, we find: 
\begin{gather}\label{eq:absorption-rates}
\begin{split}
    \frac{d \Gamma_o}{dl'} \sim & \, \frac{g_s}{V} \, , \\
    \, \frac{d \Gamma_{c}}{dl'} \sim & \,  g_s^2 \frac{l\hat{l}}{V} \, , 
\end{split}
\end{gather}
for a process with incoming strings of fixed lengths $l$ and $\hat{l}$, yielding a string of length $l'$.
Based on our findings in the previous section, namely that the typical decay renders highly excited, nonrelativistic strings, we are entitled to assume that the interactions preserve length, so $\hat{l}=l'-l$.

Again, the results can be interpreted semiclasically: absorption of an open string by an open string only occurs through the endpoints, so the rate should not depend on the lengths of the strings that interact. 
For closed strings, however, the interaction can happen anywhere along the string, so the rate is proportional to the product of the lengths. 

\subsection{Random walk arguments for other interactions}\label{sec:rws}

In the remainder of this section we use random walk arguments to conjecture the form of other interactions with strings at fixed levels but averaged or summed over states in those levels. Given the implicit interest of string physics at
high excitation, a microscopic derivation of these conjectures, as we have for 
three-point interactions above, would be valuable.

For us, these interaction rates will
play a role in formulating the Boltzmann equations for a distribution of open and closed strings, which at high level is determined by their length.
The fact that these interaction rates agree with detailed balance for the equilibrium solution of Eq.~\eqref{eq:eq-dist} is a nontrivial check.
Throughout the rest of the section, we write $d \Gamma/dl$ for all the interactions, whose particular nature should be understood by context.

\subsubsection*{High brane density or space-filling branes}

Let us begin by conjecturing the following form of the leading order hybrid interactions in the case of high D$p-$brane density, where we assume a stack of $N$ effectively overlapping parallel D$p-$branes, uniformly distributed in a transverse volume $V_{\perp}$.
Here, effectively overlapping means that the typical string is much longer than the D$p-$brane separation, and indeed the random walk argument we will use assumes that the open string is an open random walk in $d-$dimensions. 
We assume that the dimensions along the branes with volume $V_\|$ are 
effectively noncompact, and we include only effectively noncompact dimensions 
in $V_\perp$, following the discussion of Appendix \ref{sec:winding-modes}.
An example of this case is a space-filling brane ($d=p$) with only compact
dimensions transverse; then $V_\perp\equiv 1$.

For an open string closing up to yield a process open$\rightarrow$closed,
\begin{equation}
    \left.\frac{d\Gamma}{dl}\right|_{\text{ends fixed}}
\sim\frac{g_s}{l^{p/2}} \, ,
\end{equation}
if we assume that both end points are attached to the same brane. The
length-dependent factor comes from the probability for the string to close
in the dimensions along the brane. However, if we average over all incoming
strings, most of the strings will have ends attached to different branes.
For a given string attached at one end to a particular brane, the other end
can access a fraction $l^{(d-p)/2}/V_\perp$ of the $N$ branes, so the requirement
that both ends of the string are on the same brane reduces the amplitude
by a corresponding factor:
\begin{equation}
    \frac{d\Gamma}{dl}\sim \frac{V_{\perp}}{N}\frac{g_s}{l^{d/2}} \, ,
\end{equation}
where $l$ is the length of the product closed string.
In the case of space-filling branes, the entire $l^{-d/2}$ factor is the
probability of the random walk closing while all $N$ branes are accessible.
\\

Similarly, the chopping of a closed string by a D-brane to give closed$\rightarrow$open should be proportional to the length of the string since it can occur at any point, and to the density of D$p-$branes:
\begin{equation}
    \frac{d\Gamma}{dl}\sim \frac{N}{V_{\perp}}lg_s \, .
\end{equation} 
\\

Consider now the higher order absorption process (closed, open)$\rightarrow$open with lengths $(l',l-l')\rightarrow l$ within a space-filling brane.
This interaction can happen anywhere along both strings, so the dependence should be again proportional to the product of lengths:
\begin{equation}
    \frac{d\Gamma}{dl}\sim g_s^2 l' (l-l') \, .
\end{equation}
\\

The inverse emission process, however, has a probability factor $1/l'^{d/2}$ due to the requirement that a segment of length $l'$ of the string closes up.
If the initial string has length $l$, such closing up can occur along any of the $l-l'$ points.
This implies an overall probability of
\begin{equation}
    \frac{d\Gamma}{dl}\sim g_s^2\frac{l-l'}{l'^{d/2}} \, .
\end{equation}

These decay rates are almost all we need to describe strings with effectively space-filling branes. To work consistenly at order $g_s^2$
in string perturbation theory, we must also include additional 4-point interactions describing (open,open)$\rightarrow$(open,open) processes.
These are described in Appendix~\ref{sec:22openstrings}.

Note that setting $p=0$ describes the case of compact dimensions along the
branes with branes densely distributed through effectively noncompact transverse
dimensions.

\subsubsection*{Isolated stack of branes}

The random walk approach becomes particularly useful if the stack of branes is isolated, and we let the strings extend outside the $(p+1)$-dimensional worldvolume.
The closed string interactions are not modified, and leading order hybrid interactions are only modified in the case for an open string closing up: because the strings form random walks and the endpoints are confined to the $(p+1)-$dimensional braneworld, the rate is proportional to $1/l^{p/2}$.
Furthermore, since the stack is in a fixed position within the transverse directions there is no additional $V_{\perp}$.
\\

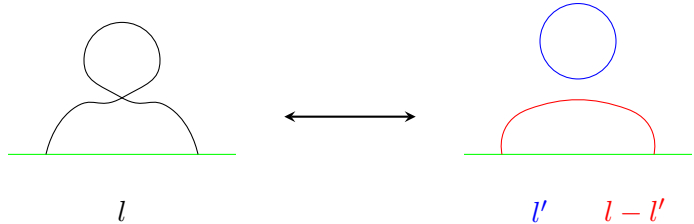
\begin{figure}
    \centering
\begin{tikzpicture}[node distance=2cm]
\tikzstyle{arrow} = [thick,<->,>=stealth]
\tikzstyle{littlearrow} = [thin, <->]
    \draw[green] (-0.5,0) -- (2.5,0);
    \draw[black]  plot[smooth, tension=1] coordinates {(0,0) (0.35,0.6) (1,0.75)   (1.5,1.25) (1,1.75) (0.5,1.25) (1,0.75) (1.65,0.6) (2,0)};
    \node (start) [xshift=3cm, yshift=0.5cm] {};
    \node (end) [right of=start] {};
    \draw [arrow] (start) -- (end);
    \draw[green] (5.5,0) -- (8.5,0);
    \draw[red]  plot[smooth, tension=1] coordinates {(6,0) (6.35,0.6) (7.65,0.6) (8,0)};
    \draw[blue] (7,1.5) circle [radius=0.5];
    \node (l1)[xshift=1cm, yshift=-0.75cm]{$\textcolor{black}{l}$};
    \node (l2) [xshift=6.5cm, yshift=-0.75cm]{$\textcolor{blue}{l'}$};
    \node (l3) [xshift=7.75cm, yshift=-0.75cm]{$\textcolor{red}{l-l'}$};
\end{tikzpicture}
    \caption{Next to leading order hybrid interactions for strings probing the directions transverse to the brane (green line).}
    \label{fig:transverse-hybrid}
\end{figure}

The second order hybrid interactions require more thought. 
Consider an open string in a process open$\rightarrow$(open,closed) given by $l\rightarrow (l-l',l')$, illustrated in Fig.~\ref{fig:transverse-hybrid}.
The total probability of the process will be given by the product of two probabilities. 
First, we must consider the probability that, chosen a point at a distance $l_x$ from an endpoint touches a point at a distance $l'+l_x$ from the same endpoint, which is given by $1/l'^{d/2}$.
Second, this point must belong to a random walk of length $l-l'$ that closes up in the directions transverse to the brane.
This is given by 
\begin{equation}
    \lr{\frac{l-l'}{l_x(l-l'-l_x)}}^{(d-p)/2}\, .
\end{equation}
Putting all of these terms together, we get (summing over all the possible points $l_x$ in which the interaction can occur):
\begin{equation}
    \frac{d\Gamma}{dl}\sim g_s^2\frac{1}{l'^{d/2}}\int_{l_c}^{l-l_c}{dl_x\, \lr{\frac{l-l'}{l_x(l-l'-l_x)}}^{(d-p)/2}} \, .
\end{equation}
where we have introduced a cutoff because our approximations break down for very small strings,\footnote{Note that the appearance of the cutoff in the upper limit is equally important, since it implies that as $l_x \to l-l_c$ there is an isolated string of length $l_c$ that is also formed.} but even if these were taken into account the length of the fundamental string would give a natural cutoff. Of course, understanding what the short strings are doing in the plasma
is a question of great interest. They include the radiation in the plasma
(which would correspond to a primordial signal in the cosmological context). We plan to  pursue this direction in the future.
\\

Consider now the inverse process of absorption of a closed string of length $l'$ by an open string of length $l-l'$. 
Similar considerations apply: the process can occur along any point within the closed string (giving a factor of $l'$), and a point separated a distance $l_x$ from the endpoint of the open string.
This point must belong to a random walk of length $l-l'$ that closes up in the transverse directions, and thus the contribution is
\begin{equation}
    \frac{d\Gamma}{dl}\sim g_s^2 \, l'\int_{l_c}^{l-l'-l_c}{dl_x\, \lr{\frac{l-l'}{l_x(l-l'-l_x)}}^{(d-p)/2}} \, .
\end{equation}
We will see that these two contributions indeed cancel in equilibrium as 
required by detailed balance.

Finally, we note that setting $p=0$ describes branes wrapped on compact
dimensions in effectively noncompact transverse dimensions. That is, the branes
are pointlike in all the dimensions we consider.

\section{The Boltzmann equations for highly excited strings}\label{sec:boltzmann}

In this section, we pose a system of Boltzmann equations for the number distribution of strings, $n_o(l)$ ($n_c(l)$) for open (closed) strings, where $n_i(l)dl$ is the number of strings in the system with length between $l$ and $l+dl$.
We will consider several increasingly complicated setups.
After a short discussion of the case studied by Lowe and Thorlacius~\cite{Lowe:1994nm} for closed strings that fill every direction (see also~\cite{Lee:1997iz} for the case with branes), we consider the case with $d$ effectively noncompact directions.
We later introduce D$p-$branes in the high and low density regimes. Since strings gravitate, there is the possibility of a Jeans instability --
as in \cite{Lowe:1994nm, Lee:1997iz} we will take the string coupling to be sufficiently small to avoid this\footnote{Our analysis
in this paper will be for Minkowski compactifications. Future studies will be in the context of an expanding universe, here the instabilility
is known of be milder and can also serve as a mechanism leading to the formation of primordial blackholes.}. 

In an ensemble of strings, there will be reactions involving strings of all lengths, so in order to look for detailed balance (and thus equilibrium solutions), we need to be careful to identify the two directions of the reaction $A\rightleftharpoons B+C$.
By doing so, using the interaction rates computed in Sec.~\ref{sec:decayrates}, we will see that detailed balance is satisfied for all channels.

Let us begin quoting the results of~\cite{Lowe:1994nm} for closed strings in zero noncompact directions.\footnote{The criterion to take dimensions compact/non-compact has been described
in the review subsection in the introduction of the article. The discussion  in the introduction was in the context of a single string. Here, the ensemble allows for arbitrary number of strings. Thus the relevant comparison is between  $(\bar{l})^{D/2}$ and the volume, $\bar{l}$ being the
length of the typical string.}
By assuming that interactions of long strings should be proportional to their length, and that interactions preserve length, the following equation was posed to describe an ensemble of highly excited strings wrapping many times the allowed volume, $V$:
\begin{equation}\label{eq:lowe-thorlacius}
\begin{split}
\frac{\partial n(l)}{\partial t}=& \frac{\kappa}{V}\left\{-\frac{1}{2}l^2 n(l)
+\frac{1}{2}\int_0^l dl'\, l'n(l')(l-l')n(l-l') \right.\\
&\left.-ln(l)\int_0^\infty dl'\, l'n(l')+\int_0^\infty dl'\, (l+l')n(l+l')\right\}\\
=&\frac{\kappa}{2V}\int_0^l{dl'\, \lr{l'n(l')(l-l')n(l-l')-ln(l)\vphantom{\frac 12}}}\\ 
 &+\frac{\kappa}{V}\int_l^\infty{dl'\, \lr{l'n(l')-(l'-l)n(l'-l)ln(l)\vphantom{\frac 12}}}\, .
\end{split}
\end{equation}
The first line is the Boltzmann equation as given in \cite{Lowe:1994nm};
we have re-written the expression after the second equality for two reasons.
The first one is that writing the interactions in this way makes the generalization easier, so let us understand the terms one by one: the first line contains interactions with strings shorter than $l$, either a fusion of strings $(l',l-l')\to l$ in the first term, or by self-decay in the second term, where all possible product strings have the same probability.
This will change in the more general setups later.
Similarly, the second line contains interactions with strings longer than $l$, either the self decay of strings of length $l'$ into $l$ and $l'-l$, or the inverse process of fusion of strings $(l,l'-l)\to l'$. 

The second reason is that writing the equation in this form makes clear how detailed balance is organised.
Detailed balance operates length by length  (with the channels organised by $l'$). 
Once we realise this, the equilibrium solution to Eq.~\eqref{eq:lowe-thorlacius} is obvious
\begin{equation}\label{eq:d0equilibrium}
    n_{eq}(l)=\frac{e^{-l/L}}{l}\, .
\end{equation}
The length by length organisation of detailed balance is an important feature, we will see that it holds in all the settings that we will study
(and this will help us to arrive at the equilibrium solutions\footnote{Recall that in~\cite{Lowe:1994nm}, the equilibrium solution was found by taking a Laplace transform.
This kind of method to solve integral equations is not available in general.}) .
\\

Henceforth, we will ignore any compact dimensions completely. If there are
effectively compact dimensions (in addition to the effectively noncompact
dimensions), every self-intersection interaction is divided by the volume
of the compact dimensions (representing the probability of the two points
intersecting in the compact dimensions), the open string joining interactions
are divided by the volume of the longitudinal compact dimensions, and the
splitting interactions are divided by the volume of the transverse compact
dimensions. As a result, we can redefine the couplings by the compact volumes
(eg, $\kappa/V_c\to\kappa$), which means we use the couplings of the 
lower-dimensional effective theory. In the Boltzmann equation from 
\cite{Lowe:1994nm} above, we can simply remove $V$ and use the coupling
$\kappa$ of the $(0+1)$-dimensional theory.

\subsection{Closed strings in arbitrary dimensions}
We are now in a position to pose Boltzmann equations in more complicated setups.
Let us consider an example of a contribution to $\partial n_c(l)/\partial t$, coming from the decay of any string of length $l'>l$.
All such contributions are summarized in the term

\begin{equation}
    \int_l^\infty{dl'\,\frac{d\Gamma_{c}}{dl}n_c(l')}\, ,
\end{equation}
which is indeed the number of product strings of length $l$ produced by a single string of length $l'$, times the number of such strings in the ensemble, and this is integrated
for all $l'>l$.

This can also be thought of as the probability that,
given a segment of length $l$ on the longer string, it closes.

With the previous comments in mind, using the worldsheet interaction rates of the previous section, we can write the Boltzmann equation in $d$ noncompact dimensions for closed strings in the absence of branes:
\begin{equation}\label{eq:boltzmannclosed}
\begin{split}
     \frac{\partial n_c(l)}{\partial t}= & 
     \frac{1}{2} \int_{l_c}^{l-l_c}{dl' \, \lr{\kappa_a\frac{n_c(l') l'\, n_c(l-l') (l-l')}{V}\, -\kappa_b ln_c(l) \lr{\frac{l}{l'(l-l')}}^{d/2}}}   \\ &
     +\int_{l+l_c}^\infty{dl' \,  \lr{\kappa_b l'n_c(l')\lr{\frac{l'}{l(l'-l)}}^{d/2}-\kappa_a \frac{ln_c(l)(l'-l)n_c(l'-l)}{V} }} \, .
\end{split}
\end{equation}
We have grouped the terms so that the first line describes the reactions $l \rightleftharpoons (l',l-l')$ involving strings shorter than $l$, and the second line contains those involving longer strings, $l' \rightleftharpoons (l,l'-l)$.
The constants $\kappa_{a,b}\sim g_s^2$ are $d$-dependent and computable but not important for the present discussion; in general $d$, they are distinct for absorption and decay due to phase space from integrals.\footnote{Note the difference from the $d=0$ case as discussed in \cite{Lowe:1994nm}.}
Note how the random walk terms appear in the same footing as the volume, suggesting an interpretation in terms of an ``effective volume" where the interaction can occur.
A similar equation as Eq.~\eqref{eq:boltzmannclosed} was written in~\cite{Copeland:1998na}, taking into account the effects of the expansion of the Universe and trading the cutoff for a length $\eta$ in the decay rates, associated with small scale structure.
Interestingly, in our case $l_c \simeq 9l_s$ (see Appendix~\ref{sec:cutoff}) which suggests that highly excited strings are random walks to a very good approximation, with points only being correlated at a distance $\sim 9l_s$.

Detailed balance now requires the same condition for both channels,
\begin{equation}
    \kappa_b V n_c(l)l^{1+d/2}=\kappa_a n_c(l')l'^{1+d/2}n_c(l-l')(l-l')^{1+d/2}\, .
\end{equation}
This has the simple solution $n_c(x)=(\kappa_b V/\kappa_a)f(x)/x^{1+d/2}$ so that $f(x)f(y)=f(x+y)$. 
The most general continuous solutions are unit and exponential functions, but the only physically reasonable ones are those that decay exponentially
\begin{equation}\label{eq:eq-dist-closed}
    n_c(l)=V \frac{\kappa_b}{\kappa_a}\frac{e^{-l/L}}{l^{1+d/2}}\, ,
\end{equation}
where $L$ is a constant to be discussed. 
The distribution resembles a Boltzmann distribution, but one must be careful in identifying $L$ directly with the temperature.
Instead, the general considerations of Eq.~\eqref{eq:noncompact-dos} at high energies require that we identify $1/L=\beta-\beta_H$.
We can nevertheless relate $L$ to an average energy density of the system, corrected by finite size effects. 
In zero dimensions, it is simply interpreted as the average length 
(which equals the average energy) of the strings, up to a constant: $L\simeq\langle l \rangle=\int_0^\infty{dl' \, l'n_c(l')}$.
In higher dimensions, it is the average energy density, corrected by the volume occupied by the string, $L\simeq\int_0^\infty{dl'\, (\kappa_a/\kappa_b V) (l'^{d/2}) l'n(l')}$.

\subsection{Open and closed strings for space-filling branes}
In this section we write the Boltzmann equation for a system with open and closed strings with space-filling or densely packed D$p$-branes:
\begin{equation}\label{eq:boltzmannbrane}
\begin{split}
     \frac{\partial n_c(l)}{\partial t}= & +\frac{b}{2N}V_\perp\frac{n_o(l)}{l^{d/2}} -a\frac{N}{V_\perp}ln_c(l) \\ &
     +\frac{1}{2} \int_{l_c}^{l-l_c}{dl' \,\lr{\kappa_a \frac{n_c(l') l'\, n_c(l-l') (l-l')}{V}\, -\kappa_b ln_c(l) \lr{\frac{l}{l'(l-l')}}^{d/2}}}   \\ &
     + \int_{l+l_c}^\infty{dl' \,  \lr{\kappa_b l'n_c(l')\lr{\frac{l'}{l(l'-l)}}^{d/2}-\kappa_a \frac{ln_c(l)(l'-l)n_c(l'-l)}{V} }} \, \\ &
    +\int_{l+l_c}^\infty{dl' \, \lr{\kappa_c \frac{(l'-l)n_o(l')}{l^{d/2}}  -\kappa_d \frac{ln_c(l)(l'-l)n_o(l'-l)}{V} }} \, .
\end{split}
\end{equation}
Here, $N$ is the number of parallel, effectively overlapping D$p$-branes in the $d-p$ transverse directions.
These span a worldvolume $V_\parallel$, and the transverse volume is denoted $V_\perp$, which together yield a total volume $V=V_\parallel \cdot V_\perp$.
Like $\kappa$, $a$ and $b$ are computable constants but are order $\mathcal{O}(g_s)$. 
We have grouped the terms so that the first line describes the leading order contributions of an open string closing up and a closed string being chopped off by a brane, the second line contains the reaction $l \rightleftharpoons (l',l-l')$, the third line contains $l' \rightleftharpoons (l,l'-l)$ as before, and the fourth line describes an open string self-intersecting to yield a pair open-closed and absorption of a closed string by an open string (with corresponding coefficients $\kappa_c,\kappa_d$). 
\\

In a similar fashion, we can write the Boltzmann equation describing the interaction of open strings at second order in string perturbation theory,
\begin{equation}\label{eq:boltzmannopenbrane}
    \begin{split}
        \frac{\partial n_o(l)}{\partial t}=&+a\frac{N}{V_\perp}ln_c(l)-\frac{b}{2N}V_\perp\frac{n_o(l)}{l^{d/2}}\\ 
        &+\int_{l_c}^{l-l_c}{dl'\, \lr{\frac{b}{2NV_\parallel}\, n_o(l')n_o(l-l')-a\frac{N}{V_\perp}n_o(l)}}\\ &
        +\int_{l+l_c}^{\infty}{dl'\, \lr{2a\frac{N}{V_\perp}n_o(l') -\frac{b}{NV_\parallel}n_o(l)n_o(l'-l)}} \\
        &+\int_{l_c}^{l-l_c}{dl' \lr{\kappa_d\frac{l'(l-l')n_c(l')n_o(l-l')}{V}-\kappa_c n_o(l)\frac{l-l'}{l'^{d/2}}}}\\
        &+\int_{l+l_c}^\infty{dl'\lr{\kappa_c\frac{n_o(l')l}{(l'-l)^{d/2}}-\kappa_d\frac{ln_o(l)(l'-l)n_c(l'-l)}{V}}}\\
        &+\text{(2-2 interactions)},
    \end{split}
\end{equation}
where the first line involves chopping off a closed string of length $l$ or an open string closing up, the second and third lines involve emission/absorption of open strings through the endpoints, the fourth and fifth line involves hybrid interactions as before, and the final line is 2-2 open string interactions that we discuss in Appendix~\ref{sec:22openstrings}.

In equilibrium, the form of $n_o(l)$ and $n_c(l)$ is fixed by detailed balance in any of the lines 
and yields
\begin{equation}\label{eq:solbraneworld}
    n_c(l)=V\frac{\kappa_b}{\kappa_a}\frac{e^{-l/L}}{l^{d/2+1}}\, , \qquad n_o(l)=\frac{2a\kappa_b N^2V_\parallel}{b\kappa_a V_\perp}e^{-l/L}
\end{equation}
and a relationship between the interaction pre-factors:
\begin{equation}
    \frac{\kappa_b}{\kappa_a}=\frac{\kappa_c}{\kappa_d}\, .
\end{equation}
Verifying this relationship from first principles would be a nontrivial check of the self-consistency of string theory.

It is worth remarking that, provided it is not zero, the string coupling does not appear in the equilibrium distributions~\eqref{eq:solbraneworld} since it cancels out in the ratios $a/b$ and $\kappa_a/\kappa_b$.
This agrees with the previous observation that our results hold to all orders in string perturbation theory.

\subsection{Open and closed strings with isolated branes}\label{sec:generalcase}

The previous setup was considered for strings moving in a $d$-dimensional 
space-filling D-brane (or, analogously, for strings so long that the separation between the branes in the transverse direction is negligible). 
In this section, we consider strings extending in the $(d-p)$-dimensional transverse volume without branes, as would occur in e.g. a toroidal compactification with a single stack of branes.

Using the interaction rates argued in Sec.~\ref{sec:decayrates}, we find
\begin{equation}\label{eq:boltzmannclosedgeneral}
\begin{split}
     \frac{\partial n_c(l)}{\partial t}= & \frac{b}{2N}\frac{n_o(l)}{l^{p/2}} -a\frac{N}{V_\perp}ln_c(l) \\ &
      +\frac{1}{2} \int_{l_c}^{l-l_c}{dl' \,\lr{\kappa_a\frac{n_c(l') l'\, n_c(l-l') (l-l')}{V}\, -\kappa_b ln_c(l) \lr{\frac{l}{l'(l-l')}}^{d/2}}}   \\ &
     + \int_{l+l_c}^\infty{dl' \,  \lr{\kappa_bl'n_c(l')\lr{\frac{l'}{l(l'-l)}}^{d/2}  -\kappa_a\frac{ln_c(l)(l'-l)n_c(l'-l)}{V} }} \, \\ &
    +\int_{l+l_c}^\infty{dl' \, \int_{l_c}^{l'-l-l_c}{  \frac{dl_x }{\lr{l_x(l'-l-l_x)}^{(d-p)/2}}}}\left(\kappa_c \frac{n_o(l')l'^{(d-p)/2}}{l^{d/2}}\right. \\ &-\left. \kappa_d\frac{l n_c(l)}{V}n_o(l'-l)(l'-l)^{(d-p)/2}\right) \, 
\end{split}
\end{equation}
and 
\\
\begin{equation}
    \begin{split}
        \frac{\partial n_o(l)}{\partial t}=&+a\frac{N}{V_\perp}ln_c(l)-\frac{b}{2N}\frac{n_o(l)}{l^{p/2}}\\
        &+\int_{l_c}^{l-l_c}{dl'\lr{\frac{b}{2NV_\parallel}n_o(l')n_o(l-l')-aNn_o(l)\lr{\frac{l}{l'(l-l')}}^{(d-p)/2}}} \\ 
        &+\int_{l+l_c}^{\infty}{dl'\, \lr{2aNn_o(l')\lr{\frac{l'}{l(l'-l)}}^{(d-p)/2} -\frac{b}{NV_\parallel }n_o(l)n_o(l'-l)}} \\
        &+\int_{l_c}^{l-l_c}dl' \int_{l_c}^{l-l'-l_c}\frac{dl_x}{\lr{l_x(l-l'-l_x)}^{(d-p)/2}}\left(\kappa_d \frac{l'n_c(l')n_o(l-l')}{V}(l-l')^{(d-p)/2}\right.  \\ & \left.-\kappa_c\frac{n_o(l)}{l'^{d/2}}l^{(d-p)/2} \right) \\
        &+\int_{l+l_c}^\infty dl'\, \int_{l_c}^{l-l_c}  \frac{dl_x }{\lr{l_x(l-l_x)}^{(d-p)/2}}\left(\kappa_c\frac{n_o(l')}{(l'-l)^{d/2}}l'^{(d-p)/2}\right. \\ 
        & \left.-\kappa_d \frac{n_o(l)(l'-l)n_c(l'-l)}{V}l^{(d-p)/2}\right) \\
        & +\text{(2-2 interactions)}\, .
    \end{split}
\end{equation}
Care is required when considering the powers of volume, but they follow from simple probability arguments.
It is immediate to show that the following distribution is an equilibrium solution
\begin{equation}\label{eq:solgeneral}
    n_c(l)=V\frac{\kappa_b}{\kappa_a}\frac{e^{-l/L}}{l^{1+d/2}}\, , \qquad n_o(l)=\frac{2a\kappa_bN^2V_\parallel}{b \kappa_a}\frac{e^{-l/L}}{l^{(d-p)/2}}\, .
\end{equation}

We study in detail the complicated contributions from 2-2 interactions in Appendix~\ref{sec:22openstrings}.

\section{Moving out of equilibrium: equilibration rates}\label{sec:perturbations}

Unlike other approaches to thermodynamics, Boltzmann equations allow us to 
study thermodynamics out of equilibrium, which has imporant applications,
for example, in cosmology. In this section, we illustrate non-equilibrium
thermodynamics of strings by considering the behaviour of fluctuations 
$\delta n (l,t)$ in the ensemble of strings, specifically finding the rate at
which these perturbations decay. 

We start by considering closed strings only because the Boltzmann equation is
simpler.
In zero noncompact dimensions, we show that the time required to reach equilibration depends on the length of the strings involved in the fluctuation, concluding that long strings equilibrate faster than short strings.
We do this by finding a family of analytic solutions that decay exponentially with a length-dependent rate $\Gamma(l)=\kappa(l^2/2+l L)/V$.
We then give a qualitative reason for this behaviour, and we conclude with similar arguments in the higher-dimensional cases.

We then turn to equilibration in the presence of D-branes, using the logic
developed with closed strings. Since splitting
and joining of open strings is lower order in string perturbation theory
than interchange interactions, equilibration of an open and closed string gas
is parametrically faster (by a factor of the string coupling) 
than equilibration of closed strings alone.
Throughout the section and in Appendix~\ref{sec:detail-perturbations}, we set $\kappa_a=\kappa_b\equiv \kappa$ for simplicity.

\subsection{Closed strings in only compact dimensions}

This section is dedicated to the study of the integro-differential equation obtained by linearizing~\eqref{eq:lowe-thorlacius} around the equilibrium solution~\eqref{eq:d0equilibrium}, which we denote $\Bar{n}(l)$.
Namely, for $n(l,t)=\Bar{n}(l)+\delta n(l,t)$, we find at order $\mathcal{O}(\delta n)$ (writing $n_c(l)$ as $n(l)$ wherever there is no room for ambiguity):
\begin{equation}\label{eq:diffeq-f}
    \frac{V}{\kappa}\frac{\partial{ \delta n(l,t)}}{\partial t}=-\lr{\frac{l^2}{2}+lL}\delta n(l,t)+\int_0^l{dl'\, l'\delta n(l',t)\lr{e^{\frac{-(l-l')}{L}}-1}}-\delta E\lr{e^{-l/L}-1}\, ,
\end{equation}
where we have and assumed finiteness of the energy of the perturbation, defined by
\begin{equation}
    \delta E\equiv \int_0^\infty {dl'\, l' \delta n(l',t)}\, .
\end{equation}
In Appendix~\ref{sec:detail-perturbations} we find two sets of solutions to this equation by converting it into a second order partial differential equation.
These have the form (up to an overall constant):
\begin{gather}
    \delta n(l,t)=e^{-l/L}/L^2\, ,\label{equildiff}\\
    \delta n(l,t)=\sqrt{\frac{\kappa}{V}}\sqrt{\frac{\pi(c+t L^2)}{2}}\frac{e^{-\frac{l}{L}+A(t)^2}}{L^2}\text{Erf}\lr{A(t),A(t)+\sqrt{\frac{\kappa}{V}}\sqrt{\frac{c+t L^2}{2}}\frac{l}{L}}\, ,\label{erfsoln}
\end{gather}
where $\text{Erf}(z_1,z_2)=\frac{2}{\sqrt{\pi}}\int_{z_1}^{z_2}{e^{-t^2}dt}$ is the incomplete error function, and
\begin{equation}
    A(t)=\sqrt{\frac{V}{\kappa}}\sqrt{\frac{c+t L^2}{2}}\lr{\frac{\kappa}{V}-\frac{1}{c+tL^2}}\, .
\end{equation}

The first solution for $\delta n(l,t)$ is simply the energy carried by strings between $l$ and $l+dl$.
This is the statement that energy is conserved in the interactions or, in the language of kinetic theory, that energy is a \textit{collisional invariant} (see e.g.~\cite{Tong:notes}). 
More interesting is the second solution, which shows the decay of a time-dependent configuration with a length-dependent decay rate given as $\Gamma (l)=\frac{\kappa}{V}\lr{\frac{l^2}{2}+l L}$ at late time, plus a time-independent term which is cancelled for zero-energy fluctuations.
\\

\begin{figure}
     \centering
     \begin{subfigure}[b]{0.45\textwidth}
         \centering
         \includegraphics[width=1.1\textwidth]{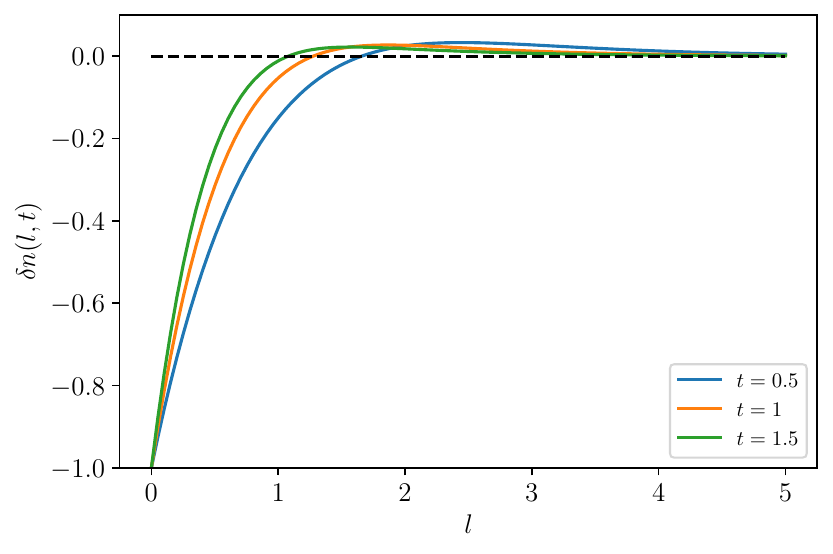}
     \end{subfigure}
     \hfill
     \begin{subfigure}[b]{0.45\textwidth}
         \centering
         \includegraphics[width=1.1\textwidth]{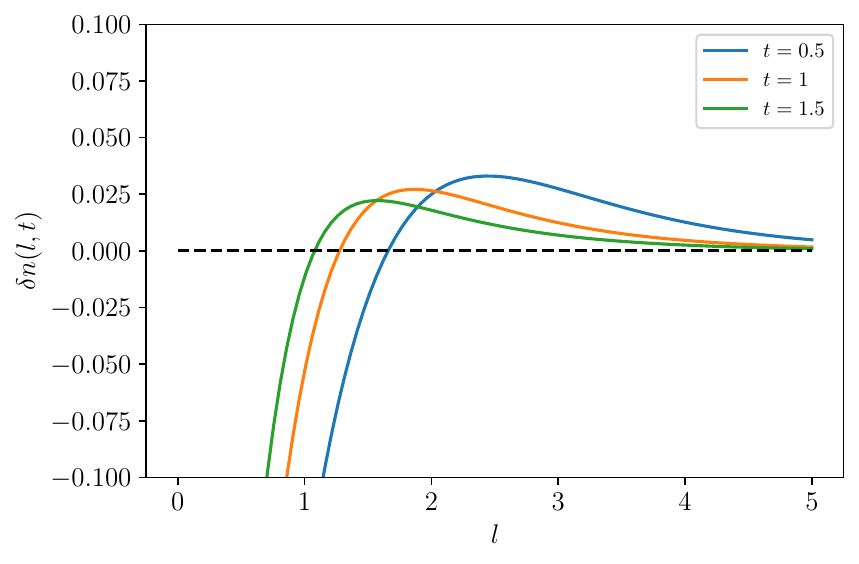}
     \end{subfigure}
        \caption{Temporal evolution of $\delta n(l,t)$ as given by~\eqref{eq:zero-e-fluctuation}, in units $L=\kappa =V=1$, and with choice of time origin $c=0$.}
        \label{fig:icfluctuation}
\end{figure}

Indeed, the only fluctuations that tend to zero at late times 
are those with $\delta E=0$; the late time form with $\delta E\neq 0$ is
proportional to (\ref{equildiff}), which is the difference between two 
equilibrium solutions $\bar n(l)$ with infinitesimally different $L$.
Meanwhile, the energy of the configurations~\eqref{equildiff} and~\eqref{erfsoln} is the same with that choice of normalization:
\begin{equation}
    \delta E=\int_0^{\infty}{dl' \, l'\delta n(l',t)}=1\, ,
\end{equation}
which is time and $c$-independent (note that $c$ determines the time origin) 
and different than zero. 
To construct a zero energy fluctuation, one must therefore add the terms so that the total energy is zero.
Thus a physically relevant fluctuation for this system is given by
\begin{equation}\label{eq:zero-e-fluctuation}
    \delta n(l,t) = \sqrt{\frac{\kappa}{V}}\sqrt{\frac{\pi(c+t L^2)}{2}}\frac{e^{-\frac{l}{L}+A(t)^2}}{L^2}\text{Erf}\lr{A(t),A(t)+\sqrt{\frac{c+t L^2}{2}}\frac{l}{L}}-\frac{e^{-l/L}}{L^2} \, ,
\end{equation}
up to an overall multiplicative constant that must be small to ensure $\delta n(l,t)\ll \Bar{n}(l)$.
Figure~\ref{fig:icfluctuation} illustrates the initial shape of the fluctuation for different times (equivalent
to different values of $c$).

\subsubsection*{Qualitative features}
Being able to obtain an analytic description of the evolution of fluctuations is a feature of the zero-dimensional case which does not carry through to more complicated situations.
However, the main features of Eq.~\eqref{erfsoln} can be argued from simple considerations that can be applied to the higher dimensional setup.

The starting point is Eq~\eqref{eq:diffeq-f}, where it is clear that zero energy perturbations satisfy
\begin{equation}\label{eq:zero-spreading}
    \frac{\partial{ \delta n(l,t)}}{\partial t}=-\frac{\kappa}{V}\lr{\frac{l^2}{2}+lL}\delta n(l,t)+\frac{\kappa}{V}\int_0^l{dl'\, l'\delta n(l',t)\lr{e^{\frac{-(l-l')}{L}}-1}}\, .
\end{equation}
There is a simple physical interpretation of this equation: the first term encodes the absorption of the fluctuation at $l$ by the bath, and the second term describes the propagation of the fluctuation in length space. While the second
term can redistribute the fluctuation, the first term should set the time
scale because it causes the fluctuation to decay everywhere.

\begin{figure} 
  \centering  \includegraphics[width=1.0\linewidth]{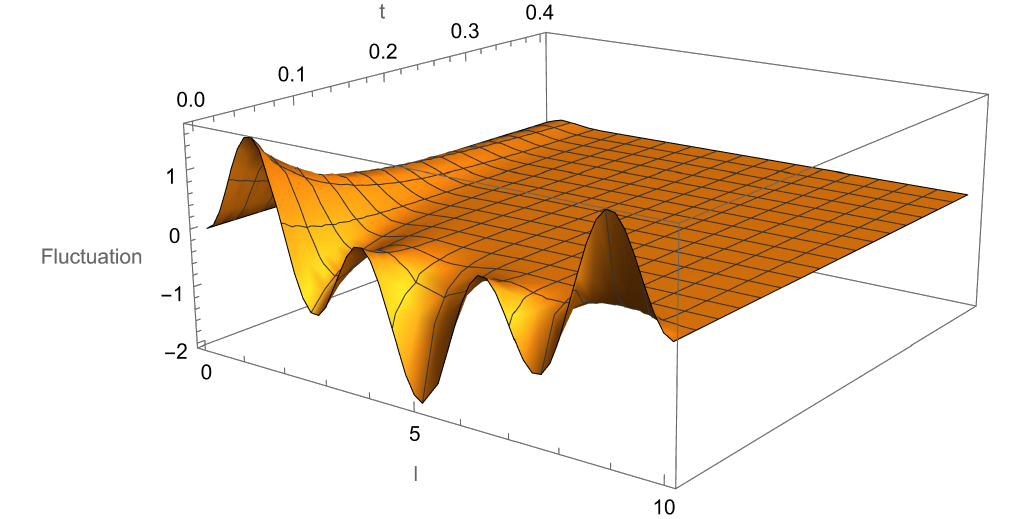}
\caption{Evolution of the fluctuation $\delta n(l,0)=\cos\lr{2\pi l/L}-\cos \lr{8 \pi l/L}$ with $L=10$ and Dirichlet boundary conditions $\delta n (0,t)=\delta n(L,t)=0$.}
\label{fig:fluctuation}
\end{figure}

We can also argue that the first term dominates as follows.
Expanding the exponential, the integral term is suppressed by powers of $(l-l')/L$, which is smaller than one except when $l\gtrsim L$, where (by consistency with $\delta n(l,t)\ll \Bar{n}(l)$) the fluctuation is exponentially suppressed.
The first term thus dominates, giving a qualitative behaviour
\begin{equation}
    \delta n(l,t)\sim \delta n (l,0)e^{-\frac{\kappa}{V}\lr{\frac{l^2}{2}+l L}t}\, ,
\end{equation}
as argued in~\cite{Lowe:1994nm}.
This qualitative behaviour is in agreement with our explicit solution, and applies for other zero energy fluctuations. 
We have indeed checked this for the initial perturbation $\delta n(l,0)=\cos\lr{2\pi l/L}-\cos \lr{8 \pi l/L}$ (with unphysical Dirichlet boundary 
conditions imposed at $l=0,L$), which carries zero energy.
The behaviour, illustrated in Fig.~\ref{fig:fluctuation}, shows that long strings tend to equilibrate faster.

\subsection{Closed strings in noncompact dimensions}
The qualitative description of the zero-energy fluctuations carried out in the previous section generalises to the higher dimensional case.
Consider now the linearization of~\eqref{eq:boltzmannclosed} around $\Bar{n}(l)=V e^{-l/L}/l^{1+d/2}$.
The evolution equation of the perturbation reads, up to $\mathcal{O}\lr{\delta n}$ terms:
\begin{equation}\label{eq:linearized-qualitative}
    \begin{split}
     \frac{\partial \lr{\delta n(l,t)}}{\partial t}= & 
     \kappa\int_{l_c}^{l-l_c}{dl'\,\lr{ (l-l')\Bar{n}(l-l')\,l'\delta n(l',t)-\frac{1}{2}\lr{\frac{l}{(l-l')l'}}^{d/2}l\delta n(l,t) }}
     \\ & 
     +\kappa \int_{l+l_c}^\infty{dl' \,  l'\delta n(l',t)\lr{\frac{l'}{l(l'-l)}}^{d/2}}-\kappa l\delta n(l,t)\frac{E}{V} \, ,
\end{split}
\end{equation}
where $E$ is the energy of the background (with density assumed to be finite), and we have neglected a term proportional to the energy of the fluctuation (which we take to be zero for simplicity).
The decaying terms require no knowledge about the shape of the fluctuation: they simply state that the fluctuation can decay through self-interactions or through interactions with the background.
We can write those contributions as:
\begin{equation}\label{closedncboltz}
     \frac{\partial \lr{\delta n(l,t)}}{\partial t}= 
     -\kappa l\delta n(l,t) \lr{\frac{E}{V}+ \frac{1}{2}\int_{l_c}^{l-l_c}{dl'\,\lr{\frac{l}{(l-l')l'}}^{d/2} }}+ \text{(spreading)} \, ,
\end{equation}
and these set the equilibration rate whenever the spreading terms are negligible, as in Eq.~\eqref{eq:zero-spreading}.
We see two contributions to the equilibration rate, which is length-dependent,
as in the $d=0$ case. The first term is linear in the length and proportional
to the energy density $E/V$; note that the energy density is given by an
incomplete gamma function that depends only on the cutoff $l_c$ and length
scale $L$. The other integral can be written as a difference of incomplete
beta functions, though each beta function diverges for even $d\geq 2$.
Keeping terms of order $l/l_c$ or greater,
\begin{equation}
\begin{split}
\Gamma_{\text{\tiny{eq}}}(l)&\simeq  \kappa\left(\frac EV l +\left\{
\begin{array}{cc} l^2/2 & (d=0)\\ (\pi/2)l^{3/2}-2\sqrt{l_c}l & (d=1)\\
l\ln(l/l_c-1) & (d=2)\\ 2ll_c^{1-d/2}/(d-2) & (d\geq 3)\end{array}\right. 
\quad\right) . \\
\end{split}
\end{equation}
This result agrees with the results at the beginning of this section for $d=0$.
For $d\geq 3$, the dominant contribution is given by the cutoff. 

We have assumed that half the terms --- those that spread the fluctuation ---
are unimportant in setting the equilibration time scale.
Those involve the spreading of the fluctuation through interactions with the background (first term in the first line in~\eqref{eq:linearized-qualitative}) or through self-decay (first term of the second line).
We cannot prove in general that these terms are always subdominant, but it is easy to see that spreading of localized fluctuations is not efficient. 
We show this in Appendix~\ref{sec:detail-perturbations}.
This argument, together with numerical results, supports the claim that the equilibration rate is given by the decaying terms. Heuristically, the reason is
that the spreading terms do not specifically move strings from overpopulated
to underpopulated values of length, which would aid in equilibration, but
rather simply change the shape of the perturbation. Hence it is the decaying
terms that set the equilibration rate.

\subsection{Open and closed strings}

Following similar considerations as before, we can study the equilibration rates of an ensemble of open and closed strings.
Since strings can split and join on D-branes at order $g_s$ in string 
perturbation theory, as opposed to the interaction of two points in the
bulk of strings at order $g_s^2$, mixed open/closed string gases can 
equilibrate parametrically faster than gases of closed strings alone.
The second order interactions for a linearized perturbation
become comparable in rate to the first order terms at best when the background
density of open strings is order $1/g_s$, in which case we should modify
the thermodynamics to include the creation and annihilation of D-branes.
As a result, we consider only the order $g_s$ interactions in this section
except for a brief comment while examining equilibration of open strings.

We will consider two types of interactions, those that transfer energy
between the open and closed string baths and those that equilibrate the open
strings. Even a perturbation that is initially entirely in the closed string
sector can equilibrate quickly by transfering energy to the open strings,
which then equilibrate rapidly. The closed strings equilibrate rapidly
with the open strings.

It seems that an open string Hagedorn phase in the early Universe could have had important consequences in reheating~\cite{Frey:2005jk}, the cosmological moduli problem and the dark radiation problem.
Similar conclusions have been reached through entropic arguments~\cite{Frey:2021jyo}. The results presented here are relevant for understanding departure
from equilibrium in such a cosmological scenario.

\subsubsection*{Transfer between open and closed strings}

Let us begin by restricting ourselves to interactions mixing the open and closed sector.
Since $a,\, b\sim g_s$ and $\kappa\sim g_s^2$, the dominant terms mixing open and closed strings read
\begin{equation}\label{eq:transfer}
    \frac{\partial \delta n_o}{\partial t}=-\frac{\partial \delta n_c}{\partial t}=a\frac{N}{V_\perp}l\delta n_c(l,t)-\frac{b}{2N}\frac{W_o(l)}{W_c (l)}\delta n_o(l,t)+\cdots\equiv A(l)\delta n_c(l,t)-B(l)\delta n_o(l,t)\, ,
\end{equation}
where we have defined $W_o(l)\equiv \text{Min}(V_\perp,l^{(d-p)/2})$, $W_c(l)\equiv \text{Min}(V,l^{d/2})$ to cover all the cases we have considered (and $\cdots$
are open-open interactions).
If we consider mixing interactions only, the number of strings involved in the fluctuation 
$\delta N(l)\equiv \delta n_o(l) + \delta n_c(l)$ is conserved as a function of
$l$ since these interactions only modify the type of string and cannot change
length. It follows that these interactions can only equilibrate the
ratio of open to closed strings at each length; at late times,
$\delta n_o(l)\to (2aN^2/bV_\perp)(W_c(l)/W_o(l))\delta n_c(l)$, which is
the same as the ratio for equilibrium distributions $\bar n_o(l)/\bar n_c(l)$.

Next, let us turn to  the equilibration rate. We want to track the evolution of
 $\Delta(l,t) \equiv \delta n_o(l,t)-\delta n_c(l,t)$. 
From the structure of the mixing terms, it is easy to see that
$\Delta$ equilibrates exponentially with relaxation rate \begin{equation}
\Gamma(l) =a\frac{N}{V_\perp}l+\frac{b}{2N}\frac{W_o(l)}{W_c (l)} .
\end{equation}
Note that interactions among open strings at
a parametrically similar rate will change $\delta N(l)$, so we cannot ignore them in a physical evolution, but  $\Gamma$ will 
still set the typical relaxation rate for $\Delta$. 

An important conclusion is that this rapid relaxation rate means that the 
whole bath of open and closed strings can equilibrate rapidly through the
interactions of the open strings.
This might have important consequences in dark radiation overproduction and the cosmological moduli problem, particularly in the context of cosmologies that have a stringy epoch of reheating
(e.g.~\cite{Frey:2005jk,Frey:2021jyo}). In particular, it would also be interesting to analyse the interplay between the present scenario and models such as those presented in~\cite{Apers:2022cyl,Conlon:2022pnx} to solve the overshoot problem, where a $\mathcal{O}(1)$ energy density in string units is explicitly realized during the kination epoch.

\subsubsection*{Open string fluctuations}
Let us now move to the study of the decay of fluctuations in an ensemble of open strings, neglecting the interactions that mix with closed strings.
Thus, at first order in perturbation theory (\textit{i.e}, neglecting 2-2 interactions), the linearized Boltzmann equation for space-filling branes or all
compact dimensions reads
\begin{equation}\label{eq:openopen}
\begin{split}
\frac{\partial \delta n_o(l,t)}{\partial t}= &\cdots +
\int_{l_c}^{l-l_c}{dl'\lr{\frac{b}{NV_\parallel } \Bar{n}_o(l')\delta n_o(l-l',t)-a\frac{N}{V_\perp}\delta n_o(l,t)}} \\
   & +2a\frac{N}{V_\perp}\int_{l+l_c}^\infty {dl'\, \delta n_o(l',t)}-\frac{b}{NV_\parallel}\bar{N}_o\delta n_o(l,t) -\frac{b}{N V_\parallel}\Bar{n}_o(l)
\int_{l_c}^\infty dl' \delta n_o(l')\, ,
\end{split}
\end{equation}
where $\bar{N}_o$ is the total number of open strings in the 
equilibrium solution (and $\cdots$ are the open-closed transfer interactions).
If, as before, we assume that the spreading terms do not contribute, we find 
\begin{equation}
    \Gamma_{\text{\tiny{eq}}}(l)=\frac{2aN}{V_\perp}\lr{\frac{l}{2}+L}\, .
\end{equation}
For all compact dimensions, this is related to the equilibration rate of 
closed strings by
\begin{equation}\label{compacteqratio}
    \frac{\Gamma_{\text{\tiny{eq,o}}}(l)}{\Gamma_{\text{\tiny{eq,cl}}}(l)}= \frac{2aN}{\kappa l}V_\parallel \, .
\end{equation}
(The power of $l$ in the denominator is lower if any dimensions are 
non-compact.)

With a single stack of branes, the linearized Boltzmann equation is
\begin{equation}
\begin{split}
\frac{\partial \delta n_o(l,t)}{\partial t}=& 
\int_{l_c}^{l-l_c}{dl'\left(\frac{b}{NV_\parallel } \Bar{n}_o(l')\delta n_o(l-l',t)
-aN \delta n_o(l)\left(\frac{l}{l'(l-l')}\right)^{(d-p)/2}\right)}\\
&+2aN\int_{l+l_c}^\infty {dl'\, \delta n_o(l',t)\left(\frac{l'}{l(l'-l)}\right)^{(d-p)/2}}-\frac{b}{NV_\parallel}\bar{N}_o\delta n_o(l,t) -\frac{b}{N V_\parallel}\Bar{n}_o(l)
\int_{l_c}^\infty dl' \delta n_o(l')\, .
\end{split}
\end{equation}
In this case, the equilibration rate is
\begin{equation}
    \Gamma_{\text{\tiny{eq}}}(l)=\frac{b}{N}\frac{\bar{N}_o}{V_\parallel}+aN
\int_{l_c}^{l-l_c}dl' \left(\frac{l}{l'(l-l')}\right)^{(d-p)/2} ,
\end{equation}
where the open string density is given by an incomplete gamma function
and we recognize the second term as the integral in (\ref{closedncboltz}).
The ratio with the closed string equilibration rate is similar to that in
(\ref{compacteqratio}), except the incomplete gamma and beta functions
the open strings have lower effective dimensionality, which actually increases
the numerator.

It follows that open strings equilibrate much faster than closed strings for spacetime-filling or isolated branes (except for very long strings in $d<3$ 
noncompact dimensions). This justifies the picture we described at the start 
of this subsection: the open strings equilibrate quickly, and the interactions
mixing open and closed strings transfers that equilibrium to the closed string
sector. It is also worth noting that the ratio of equilibration rates
for the open string splitting and joining interactions to the open string
$2-2$ interactions is parametrically the same as (\ref{compacteqratio})
in all compact dimensions, following from the form of the interactions
given in appendix \ref{sec:22openstrings}.

\subsubsection*{An exact partial solution}

If all the dimensions are compact, the open string Boltzmann equation
at leading order in string perturbation theory is the combination of
equations (\ref{eq:transfer},\ref{eq:openopen}), as in \cite{Lee:1997iz}:
\begin{equation}\begin{split}
\frac{\partial n_o}{\partial t} =& \frac{aN}{V_\perp} ln_c(l) -
\frac{b}{2NV_\|} n_o(l)+\int_0^ldl'\left[\frac{b}{2NV_\|}n_o(l')n_o(l-l')
-\frac{aN}{V_\perp}n_o(l)\right]\\ 
&+\int_l^\infty dl'\left[\frac{2aN}{V_\perp}n_o(l') 
-\frac{b}{NV_\|}n_o(l)n_o(l'-l)\right] .
\end{split}\end{equation}
Integrating this equation over all lengths yields 
\begin{equation}
\frac{\partial N_o}{\partial t} = \frac{aN}{V_\perp}(E_c+E_o)-\frac{b}{2NV_\|}
(N_o+N_o^2)= \frac{b}{2NV_\|}\left(\bar{N}_o+\bar{N}_o^2-N_o-N_o^2\right)
\end{equation}
for the total number of open strings $N_o(t)$, where $E_c+E_o$ is the total
conserved energy and $\bar{N}_o$ is the equilibrium value.

Depending on whether the initial value is less or greater than the equilibrium
value, the number of open strings is
\begin{equation}\begin{split}
N_o(t) =& \left(\bar{N}_o+\frac 12\right)\tanh\left[\frac{b}{2NV_\|}
\left(\bar{N}_o+\frac 12\right)(t-t_0)\right]-\frac 12\quad\text{or}\\
N_o(t) =&\left(\bar{N}_o+\frac 12\right)\coth\left[\frac{b}{2NV_\|}
\left(\bar{N}_o+\frac 12\right)(t+t_0)\right]-\frac 12 .
\end{split}\end{equation}
In either case, the solution at late times is an exponentially decaying 
approach to equilibrium with a time scale approximately 
$b\bar{N}_o/NV_\| = 2aNL/V_\perp$. That is roughly the relaxation rate
at the typical length scale $L$
for either energy transfer between the open and closed strings or 
equilibration among open strings. The number $N_o$ of strings of course only 
carries a small amount of information about the equilibration of the whole
distribution of strings; however, it should relax with about the same time
scale, so this solution is a useful check on our above results.

\section{Conclusions}\label{sec:conclusions}

In this paper, we have generalized the Boltzmann equation approach for perturbative string theory in a flat background.
Let us summarize our results:
\begin{itemize}
    \item We have started in Sec.~\ref{sec:general-argument} by describing the phase space of string theory in a toroidal background and, through a simple, background independent thermodynamic argument, found a simple equilibrium solution that confirms the free string computations.
    This equilibrium solution has been used as a proxy for the setup we have considered in the rest of the paper.
    \item In Sec.~\ref{sec:decayrates}, we have computed decay and absorption rates for typical strings at leading order in string perturbation theory, folowing~\cite{Manes:2001cs}.
    In particular, the form of the decay rate allowed us to argue that the thermodynamics is governed by long, nonrelativistic strings with winding and momentum modes which, for highly excited strings, give a negligible contribution to the total energy of the string.
    Furthermore, the shape of the interaction rates can be analyzed through simple random walk arguments that led us to conjecture the form of other interactions, which for technical reasons are very hard to compute. As we see later, these conjectured interaction rates pass a nontrivial consistency check of satisfying detailed balance, so a worldsheet derivation would be very interesting.
\item  In Sec.~\ref{sec:boltzmann}, we have posed a system of        Boltzmann equations describing three different setups: first, we have considered a compactification with no D-branes and $d$ effectively noncompact dimensions.
    Second, we introduced space-filling D-branes in the framework. The same setup applies for parallel D$p$-branes at sufficiently high energies, where the open strings are so long compared to the typical brane separation that they probe the whole transverse space. This case should also be similar, for instance, to strings at the tip of a Klebanov-Strassler throat, used in string compactifications to create hierarchies among energy scales~\cite{Klebanov:2000hb,Giddings:2001yu}.
    Third, we have allowed the open strings to probe the transverse directions (or, analogously, reduced the energy so that the typical string is shorter than the interbrane separation).
    All of these equations admit equilibrium solutions compatible with our general argument. 
    This shows the validity of the approximation of describing the typical string by its length and of our conjectured form for decay rates.
\item Lastly, we have given a taste of out-of-equilibrium physics in Sec.~\ref{sec:perturbations} by studying the behaviour of an ensemble of closed strings under zero-energy perturbations, which are to be interpreted as fluctuations.
    We have obtained an analytic expression in Eq.~\eqref{eq:zero-e-fluctuation} for such fluctuations, and computed its equilibration rate.
    The intuition we developed in the $d=0$ case in absence of branes has allowed us to infer the behavior of equilibration rates in more complicated situations.
    In particular, we argued that, in a situation where open and closed strings are present, equilibration proceeds rapidly in the open string sector, and closed strings equilibrate via rapid interactions with the open strings.
    This result agrees with equilibrium notions but provides a dynamical mechanism for the approach to equilibrium.
\end{itemize}

\subsection*{Caveats and future directions}
The main goal of this paper is to develop the technology of the Boltzmann equation in string thermodynamics and illustrate its power.
Before concluding, let us comment on caveats of this approach and issues that it cannot address, which open potential future research directions. \\

One issue that cannot be addressed by the Boltzmann equation approach is the limitations of the canonical ensemble description, as discussed extensively in~\citep{Abel:1999rq}.
A way to understand the issue is to compute the energy density contained in the equilibrium distributions that we find for a gas of closed strings,
\begin{equation}
E \sim \int_{l_c}^{\infty}{dl\,  l \frac{e^{-l/L}}{l^{1+d/2}}} \, ,
\end{equation} 
which is finite for $d > 2$, even as $L \to \infty$, due to the monotonically decreasing function that multiplies the exponential term of the density of states.
This indicates that the onset of the Hagedorn transition requires a finite energy density, and the canonical ensemble cannot describe energy densities larger than these ones.
Perhaps the inclusion of a background of infinitely long strings, as indicated by~\cite{Copeland:1998na}, could be used to study this regime within a Boltzmann equation formalism, but this is beyond the scope of the present work.
Happily, however, in the presence of open strings in $d-p\leq 4$ effectively noncompact transverse directions (or on spacefilling branes), the equilibrium distributions we find are valid for any range of energy densities.
The open string dominance in the thermal ensemble was well known~\citep{Lee:1997iz,Abel:1999rq}, and a relevant addition of our paper is the computation of the rate of energy transfer to open string fields in Section~\ref{sec:perturbations}. \\

Two other difficulties are due to the action of gravity, which is always present in string theory.
First, we have assumed homogeneity and isotropy, so the interaction rates, couplings, and number distributions are not spacetime dependent.
However, gravitational backreaction necessarily generates inhomogeneities of the temperature (or $L$ in our formalism) in space and time; in the Euclidean description, the radius of the thermal circle is not stabilized.\footnote{If the dilaton is not stabilized, we expect parametrically similar considerations to apply to the spacetime variation of the string coupling.}
The question is whether there is an approximate local notion of equilibrium in string thermodynamics, which depends on whether the strings can equilibrate faster than density perturbations grow. The latter time scale is given by the Jeans instability in flat spacetime with a rate $\Gamma_J\sim\sqrt{G_N \rho}\sim g_s \sqrt{\rho}$ (the Hubble rate in cosmology is parametrically similar) with $\rho$ the energy density.
In comparison, we consider, for example, open and closed strings on a brane that fills the noncompact directions, the equilibration rate is parametrically $\Gamma_o\sim g_s N_o/V_\|\sim g_s L$ (ignoring length dependence). The density is $\rho=E_o/V_\|\sim L^2$, so $\Gamma_o\sim\Gamma_J$. Whether a typical string maintains equilibrium is therefore sensitive to details of the compactification model, and we leave a more detailed investigation for future work. (Closed strings alone equilibrate more slowly and are unlikely to remain in equilibrium except for the longest strings.)
In space, the Jeans length, or the length scale of growing density perturbations, is $l_J\sim 1/\sqrt{G_N\rho}$; a local concept of equilibrium requires that $l_J$ be larger than the excursion of a typical string, $L^{1/2}$, which requires $L^{3/2}\lesssim 1/g_s$ for the same open string case. However, we should already require $\rho\sim L^2\lesssim 1/g_s$, or else D-branes should appear in the thermal ensemble, so the string gas has an approximate notion of local thermal equilibrium.
It is also worth noting that the Jeans instability drives black hole formation in the gas of strings, so the above comparison also determines whether an equilibrium Hagedorn phase can exist before a transition to an inhomogeneous phase of black holes and radiation.\\

Another gravitational effect we do not address is self-gravity of the individual strings, as in~\cite{Horowitz:1997jc}.
Such interactions are only important in $d \leq 3$ for a range of couplings, so our setup should work below such coupling.
Nonetheless, a way to address this issue within a Boltzmann equation approach would be studying the interaction rates in the Horowitz-Polchinski (HP) background, and this should yield $1 \to 2$ and $2 \to 1$ interaction rates of random walks whose shape is determined by a long-range self-interaction.
Another way to figure out the form of the interaction rates would be to compute the density of states in equilibrium and apply inverse logic in the sense we have used in this paper, namely answer to the question ``which interaction rates allow for this equilibrium distribution imposing detailed balance?" \\

Many future directions are now open -- both of  phenomenological and formal character.
The computation of the gravitational wave background that would arise as a consequence of strings in the phases we have studied is under progress~\cite{Frey:2024in}.
Other interesting directions include the  extension of our results to more complicated backgrounds (such as warped throats, or the HP solution), explicit realizations of models of string gas cosmology~\cite{Brandenberger:1988aj, Brandenberger:2006vv, Brandenberger:2006xi, Nayeri:2005ck}, a recent variation \cite{Melcher:2023kpd}, and cosmologies
with epochs with of stringy reheating~\cite{Frey:2005jk, DiMarco:2019czi,Apers:2022cyl,Conlon:2022pnx,Gubser:2003vk,Barnaby:2004gg,Kofman:2005yz}
(see~\cite{Cicoli:2023opf, Brandenberger:2023ver} for recent reviews on string cosmology).
Our results when appropriately generalized  should be useful  for the cosmic string community. Finally, it would be interesting to carry out worldsheet computations to verify conjectures about scattering processes we have made based on the random walk picture of highly excited strings.

\section*{Acknowledgements}

We would like to thank Shanta de Alwis, Sebastián Céspedes, Ed Copeland, Amrita Ghosh, Tuhin Ghosh, Chris Hughes, Rishi Mouland, Mario Ramos Hamud, Lárus Thorlacius, David Tong and Yoav Zigdon for stimulating discussions.
FM is funded by a UKRI/EPSRC Stephen Hawking fellowship, grant reference EP/T017279/1, partially supported by the STFC consolidated grant ST/P000681/1 and funded by a G-Research grant for postdocs in quantitative fields.
 The work of FQ has been partially supported by STFC consolidated grants ST/P000681/1, ST/T000694/1. The work of GV has been partially supported by STFC consolidated grant ST/T000694/1. The work of ARF was supported by the Natural Sciences and Engineering Research Council of Canada Discovery Grant program, grant 2020-00054. 
The work of RM has been supported by a fellowship grant from INFN projects ST\&FI ``String Theory And Fundamental Interactions'' and GAST ``Gauge And String Theories''.

\appendix
\section{Decay rates with winding modes}\label{sec:winding-modes}

In this appendix we study the more complicated case of the computation of the decay rate for the typical highly excited string in the case where it can wind around the compact directions.
We will only consider closed strings, and give a short discussion of how this applies to open strings in the end.

Since, as argued in the main text, the amplitude for the process is the same at fixed initial winding (which makes sense, since this is a local process), we start from the expression analogous to Eq.~\eqref{eq:particular-rate}:
\begin{equation}
    \Gamma_{c}\lr{M \rightarrow M_1+M_2}\simeq \frac{g_s^2}{M^2}\mathcal{N}_L\mathcal{\Tilde{G}}_L\mathcal{N}_R\mathcal{\Tilde{G}}_Rk^{d-2} \, ,
\end{equation}
where we have defined
\begin{gather*}
    \mathcal{N}_{a} \equiv (N_{a}+1-N_{a,1}-N_{a,2})\, , \\
    \mathcal{\Tilde{G}}_a \equiv  \frac{\mathcal{G}(N_{a,1})\mathcal{G}(N_{a,2})}{\mathcal{G}(N_a)}\sim \lr{\frac{N_{a,1}N_{a,2}}{N_a}}^{-(D+1)/4}e^{-(\sqrt{N_a/\alpha '}-\sqrt{N_{1,a}/\alpha '}-\sqrt{N_{2,a}/\alpha '})/T_H}
\end{gather*}
(for $a=L,R$).
We are interested in fixing the initial conditions for the initial string and the mass of one of the second strings.
Then we will integrate over all possible winding and KK modes of both strings (which in the case of closed strings are related by conservation equations), and all possible masses $M_2$ that are compatible with $M_1$ to find the total production rate of $M_1$, as in the main text.
Let us introduce some useful notation first, following~\cite{Chen:2005ra}.
Define the vectors $Q_{i,\pm}$ through their $d_c$ components:
\begin{equation}
    Q^{(j)}_{i,\pm}\equiv {\frac{n_i^{(j)}}{R_j}\pm \frac{\omega_i^{(j)} R_j}{\alpha'}}\, ,
\end{equation}
where $j$ labels compact directions and $i$ labels the string in consideration (hence have $Q_\pm, Q_{1,\pm}$ and $Q_{2,\pm}$).
The mass of the strings and conservation of KK and winding read
\begin{gather*}
        M_i^2=\frac{4}{\alpha'}(N_{i,L}-1)+Q_{i,-}^2=\frac{4}{\alpha '}(N_{i,R}-1)+Q_{i,+}^2\, ,\\
        Q_{\pm}^{(j)}=Q_{\pm,1}^{(j)}+Q_{\pm,2}^{(j)}\, ,
\end{gather*}
where $Q_{i,\pm}^2\equiv \sum_{j=1}^{d_c} \lr{Q_{i,\pm}^{(j)}}^2$ indicates the magnitude of the vector.
As in the main text, the important physics is in the exponentials.
The least suppressed decay rate is that satisfying $\sqrt{N_1}+\sqrt{N_2}=\sqrt{N}$, and thus converting energy from oscillators into KK or winding modes or kinetic energy is exponentially penalized.
It is easy to show that, in terms of $Q_{\pm}$ (and setting $N_i-1\simeq N_i$ for simplicity), the difference is minimized for 
\begin{equation}
    \frac{\sqrt{Q_{\pm,1}^2}}{M_1}=\frac{\sqrt{Q_{\pm}^2}}{M_1+M_2}\, ,
\end{equation}
and 6 introducing this expression in the exponential, its argument is minimized for $M_1+M_2=M$, \textit{i.e}, a negligible noncompact kinetic energy, as before.
It follows that the typical string has an energy in KK and winding modes proportional to its length and is very nonrelativistic in the noncompact directions.

With this knowledge, we can go ahead and study the total production of strings of mass $M_1$.
Defining\footnote{We depart here from the notation in~\cite{Chen:2005ra} because this choice of $\alpha$, as opposed to $\alpha =M/Q$ simplifies the exposition.} a normalized vector $\alpha_{\pm}\equiv Q_{\pm}/M$, we can parameterize deviations of $Q_2/M_2$ from $\alpha$ through $Q_{2,\pm}/M_2\equiv \alpha_{\pm}+\varepsilon_{\pm}$, with small components $\varepsilon_{\pm}^{(j)}\ll 1$ due to the exponential supresion. 
Expanding for small $\varepsilon_{\pm}^2\equiv \sum_{j=1}^{d_c}{\lr{\varepsilon_\pm^{(j)}}^2}$ and $k^2$, we find:

\begin{equation}
    \Tilde{\mathcal{G}}_L\sim \exp{\lr{-\frac{k^2}{4 T_H\sqrt{1-\alpha_{-}^2}}\frac{M}{M_1M_2}}}\exp{\lr{-\frac{\varepsilon_{-}^2}{4T_H(1-\alpha^2_{-})^{3/2}}\frac{M M_2}{M_1}}}\, ,
\end{equation}
where we have written $M_1+M_2\simeq M$.
The exact same discussion applies to $\Tilde{\mathcal{G}}_R$.
If we are interested in the total production of strings of mass $M_1$ from strings with mass $M$ and quantum numbers $Q_{\pm}$, we need to integrate over all possible $k$ and $Q_{\pm}$:
\begin{equation}
    \Gamma_{M_1}\sim  \int_0^{M-M_1}{dt\, M_2  
\int_{\pm ,\, M_2}{dQ_{+,2}dQ_{-,2}\,  \frac{g_s^2}{M^2}\mathcal{N}_L\Tilde{\mathcal{G}}_L\mathcal{N}_R\Tilde{\mathcal{G}}_R k^{d_{nc}-2}}}\, ,
\end{equation}
where the limits of integration in $k$ have been chosen so that the levels are positive, the subscript $\pm , \, M_2$ indicates two integrations in a $d_c$ dimensional sphere of radius $M_2$ (recall that every $Q_a$ integration stands for $d_c$ integrations in $dQ_a^{(j)}$).
The appropriate variables of integration are $\lbrace k, \,  \varepsilon^{(j)}\rbrace$, and it is important to notice that $dQ_{\pm,2}^{(j)}=M_2d\varepsilon_\pm^{(j)}$.
This last change of variable is crucial, since it carries a $M_2^{2d_c}$ factor.
We find:
\begin{equation}
    \Gamma_{M_1}\sim \frac{g_s^2}{M^2}M_2^{2d_c}\int_0^{M-M_1}{dt\, M_2 \int_{\varepsilon_\pm+\alpha_\pm \leq 1}{\varepsilon_+^{d_c-1}d\varepsilon_{+}\, \varepsilon_{-}^{d_c-1}d\varepsilon_{-}\mathcal{N}_L\Tilde{\mathcal{G}}_L\mathcal{N}_R\Tilde{\mathcal{G}}_R k^{d_{nc} -2}}}\, .
\end{equation}
We are almost there. 
The last step is to convert the integrals in numbers through a change of variables, 
\begin{equation}
        y_\pm^2\equiv \frac{\lr{\varepsilon_{\pm}}^2}{4T_H(1-\alpha^2_{\pm})^{3/2}}\frac{M M_2}{M_1}\, .
\end{equation}
The integrals in $y$ yield factors of $\lr{M_1/M M_2}^{d_c}$.
Combined with the $M_2^{2d_c}$ obtained from the change of variables $Q\to \varepsilon$, this yields $\lr{M_1M_2/M}^{d_c}$, which cancels the $d_c$ contribution from the oscillator modes, $\lr{M_1M_2/M}^{-d-d_c-2}$.
The logic then follows as in the main text with the $t$ integral, yielding Eq.~\eqref{eq:decay-rates} with $d\to d_{nc}$ and additional factors arising from the different integrations.
\section{2-2 open string interactions}\label{sec:22openstrings}

\subsection{Strings in the worldvolume}

In this appendix we argue for the form of 2-2 interactions among open strings in both of the cases studied in the main text. 
We make explicit the detailed balance and argue why our treatment for space-filling branes (or densely packed branes) is identical to~\cite{Lee:1997iz}, who posed the system in a different way. Note that all these interaction rates appear with a power of $1/V=1/(V_\| V_\perp)$ in the Boltzmann equations, and with $\mathcal{O}(g_s^2)$ pre-factors, which we ignore for simplicity.

Let us disentangle this. 
The structure of the interactions is $A+B\leftrightarrow C+D$, with $A=x$, $B=y$, $C=l$, $D=x+y-l$.
We study this process as follows.
The positive contributions to $\partial n(l)/\partial t$ arise from the ($\rightarrow$) process, and we fix $A$, $B$ and $C$, which in turn fixes $D$.
For the inverse ($\leftarrow$) process, we fix the product strings $A$ and $B$, a mother $C$, and then sum over all points of the string $D$ (of fixed length once $A$, $B$ and $C$ are known) that can yield this process. 
This is precisely, for any point and fixed $A$, $B$ and $C$, the inverse channel.
\begin{itemize}
    \item Two short strings.
    Consider a string of length $x<l$ exchanging modes with a string of length $y$ so that $l-x<y<l$.
    This process yields a pair with lengths $l$ and $x+y-l$.
    With the D$p$-brane case in mind, let us argue for a general setup, counting every point in which the interaction can happen.
    For small incoming strings, we can only possibly choose a segment of $x$ of length $l_x$ between $0$ and $x+y-l$, since otherwise the result of the interaction is a pair of strings shorter than $l$.
    We thus have
    \begin{equation}
        \int_0^l{dx\, n(x) \int_{l-x}^{l}{dy\, n(y) \int_0^{x+y-l}{dl_x}}}\, 
    \end{equation}
    In order to see how the cancellation occurs channel by channel, we study the absorption of a string of length $l$ by a string of length $x+y-l$ to yield strings of fixed length $x$ and $y$, over which we later sum, and consider all the points of the string of length $x+y-l$ which collides with $l$ that can yield a pair of such strings (which are, in this case, all the points). 
    Thus, the form of the interaction satisfies
    \begin{equation}\label{eq:short2-2}
        -\int_0^l{dx\, \int_{l-x}^l{dy\, \int_0^{x+y-l}{d\Tilde{l}\, n(l)n(x+y-l)}}}\, ,
    \end{equation}
    And it is immediate to note that, because the equilibrium solution is an exponential, $n(x+y)=n(x)n(y)$ and the cancellation occurs.

    The last integration is in this case immediate because, provided the points collide, the interaction occurs with the same probability for every point.
    This will change when the strings are free to probe the volume transverse to the brane, and the interaction point on each string is weighted by
    the probability that it reaches the interaction point in the transverse spatial dimensions (given that each string is a closed loop in the transverse dimensions).

    \item Two long strings.
    The analysis follows similarly: we fix $x$ and $y$ and sum over all possible points of a string of length $x+y-l$ that yield the process $(l,x+y-l)\rightarrow (x,y)$, or viceversa.
    We find
    \begin{equation}
    \begin{split}
       & \int_l^{\infty}{dx\, n(x)\int_l^{\infty}{dy\, n(y) \int_0^l{dl_x}}} \\
       & -\int_l^{\infty}{dx\, \int_l^{\infty}{dy \, \int_0^l{d\Tilde{l}\, n(l)n(x+y-l)}}}\, .
    \end{split}
    \end{equation}
    The cancellation at equilibrium follows similarly.
    \item Mixed interactions.
    The remaining terms yield the mixed interactions:
    \begin{equation}
    \begin{split}
        & 2\int_0^l{dx\, n(x) \int_l^{\infty}{dy\, n(y) \, \int_0^x{dl_x}}} \\
        & -2\int_0^l{dx\, \int_l^{\infty}{dy\, \int_0^x{d\Tilde{l}\, n(l) n(x+y-l)}}} \, .
        \end{split}
    \end{equation}
\end{itemize}

The positive terms are manifestly the same as the positive terms of \cite{Lee:1997iz}.
One might be concerned that the sum of the negative terms do not obviously add up to the negative contribution argued in~\cite{Lee:1997iz}.
In their setup, they fix $C$ and $D$, in the spirit that we need to sum over all possible strings that can interact with our string of length $l$, and for fixed $C$ and $D$ sums over all possible points of both strings where the interaction can occur.
The term they write is the following:
\begin{equation}\label{eq:2-2thorl}
    -ln(l)\int_0^\infty{dl' \, l'n(l')}=-n(l)\int_0^\infty{dl' \, n(l')\int_0^{l'}{d\Tilde{l} }}\int_0^l{dx}\, .
\end{equation}
Where the second equality will simplify our derivation.
Let show that the terms are equivalent indeed.
Assuming convergence of $n(x)$, and changing variables $y\rightarrow x+y-l$, the short-short interaction of Eq.~\eqref{eq:short2-2} can be re-expressed as:
\begin{equation}
\begin{split}
    -\int_0^l{dx\, \int_{l-x}^l{dy\, \int_0^{x+y-l}{d\Tilde{l}\, n(l)n(x+y-l)}}}=&-n(l)\int_0^l{dx}\int_0^\infty{dy \, n(y)\int_0^y{dz}}  \\ 
    &+n(l)\int_0^l{dx}\int_x^\infty{dy\, n(y)\int_0^y{dz}}\, .
\end{split}
\end{equation}
The first term is precisely Eq.~\eqref{eq:2-2thorl}, and thus all what remains is to show that the following term vanishes:
\begin{equation}\label{eq:2-2-vanish}
    \int_0^l{dx}\int_x^\infty{dy\, n(y)\int_0^y{dz}}\, -\int_l^{\infty}{dx\, \int_x^{\infty}{dy \,n(y) \int_0^l{d\Tilde{l}\, }}}-2\int_0^l{dx\, \int_x^{\infty}{dy\,n(y) \int_0^x{d\Tilde{l}\,}}} \, .
\end{equation}
where we have changed variables as before in all the contributions.
It is tedious (though possible) to show directly that (\ref{eq:2-2-vanish})
vanishes, but the following derivation is shorter.
First, we note that this quantity vanishes trivially for $l=0$.
Next, consider the derivative with respect to $l$, which is
\begin{equation}
\int_l^\infty{dy\, y \, n(y)}-l\int_l^\infty{dy\, n(y)}-\int_l^\infty{dx\, \int_x^\infty{dy\, n(y)}}\, .
\end{equation}
To see that this vanishes identically, we can write all the terms as double
integrals:
\begin{equation}\left(\int_l^\infty dy\int_0^y dx-\int_l^\infty dy\int_0^l dx
-\int_l^\infty dx\int_x^\infty dy\right) n(y) .\end{equation}
Assuming convergence, we can reverse the order of integration on the last term,
so we see that the first term precisely cancels the last two terms.
Since the derivative vanishes identically and the quantity of interest vanishes
at $l=0$, it must vanish everywhere.
Thus that our approach and that of~\cite{Lee:1997iz} are equivalent.

Our approach is however more useful for two reasons.
First, it makes detailed balance obvious.
Second, it allows for a generalization to the case where the strings are not confined in the worldvolume, where the different points have different probabilities of interaction.
We now turn to this case.

\subsection{Open strings probing the transverse directions}

In this section we conjecture the form of the interactions based on random walk arguments, and show that indeed the equilibrium distribution~\eqref{eq:solgeneral} satisfies them.
The key observation is that for two colliding points at a distance $l_x$ ($l_y$) from and endpoint of a string of length $x$ ($y$), we must impose the probability that both endpoints touch the brane, \textit{i.e:} that the strings form closed random walks through the transverse directions. 
Again, note that all these interaction rates appear with a power of $1/V_\|$ in the Boltzmann equations.

As we have seen in Sec.~\ref{sec:decayrates}, the probability of one point at a distance $l_x$ from the endpoint of a string of length $x$ to lie in a particular point in the transverse directions to a D$p$-brane is given by $(x/(l_x(x-l_x))^{(d-p)/2}$ in a $d+1$-dimensional spacetime.
Thus, the probability for two such points to collide is given by 
\begin{equation}
    A(x,y,l_x,l_y) := \lr{\frac{xy}{l_x(x-l_x)l_y(y-l_y)}}^{(d-p)/2}\, .
\end{equation}
The interaction rates must take this into account, so we find for the short-short channel
\begin{equation}
\begin{split}
    &\int_0^l{dxn(x)\int_{l-x}^l{dy n(y)\int_0^{x+y-l}{dl_x A(x,y,l_x,l+l_x-x)}}}\\
    &-\int_0^l{dx\, \int_{l-x}^l{dy\, \int_0^{x+y-l}{d\Tilde{l}\, n(l)n(x+y-l)A(x+y-l,l,\Tilde{l},x-\Tilde{l})}}}\, ,
\end{split}
\end{equation}
where we have used conservation of length.
The key observation is that, using Eq.~\eqref{eq:solgeneral}, the numerator $1/x^{(d-p)/2}$ is cancelled for both distributions, yielding the product of exponentials as before, and the denominators in $A(x,y,l_x,l_y)$ are the same function in both lines.
Detailed balance is thus satisfied.

Similarly, for the long-long channel
\begin{equation}
    \begin{split}
    &\int_l^\infty{dx \, n(x)\int_l^\infty{dy \, n(y) \int_0^l{dl_x \, A(x,y,l_x,l-l_x)}}} \\
    &-\int_l^{\infty}{dx\, \int_l^{\infty}{dy \, \int_0^l{d\Tilde{l}\, n(l)n(x+y-l)A(l,x+y-l,\Tilde{l},x-\Tilde{l})}}}\, 
    \end{split}
\end{equation}
And lastly for the mixed channel
\begin{equation}
    \begin{split}
        &2\int_0^l{dx\, n(x) \int_l^\infty{dy \, n(y) \int_0^x{dl_x\, A(x,y,l_x,l-l_x)}}} \\
        &-2\int_0^{l}{dx\, \int_l^{\infty}{dy \, \int_0^x{d\Tilde{l}\, n(l)n(x+y-l)A(l,x+y-l,\Tilde{l},x-\Tilde{l})}}} \, .
    \end{split}
\end{equation}
Following the same philosophy, the contributions cancel out with the solution~\eqref{eq:solgeneral}.

\section{Cutoff dependence and equilibration rates}\label{sec:cutoff}
In this appendix we study the divergence structure of the decay rates and argue for its origin and a reasonable value for the cutoff $l_c$.
Recall that, for closed strings decaying in $d$ effectively noncompact directions in a process $l \to (l',l-l')$, the decay rate diverges as
\begin{equation}\label{eq:divergent-decay}
\frac{d \Gamma}{dl} \sim \frac{1}{(l'(l-l'))^{d/2}} \, .
\end{equation} 
and thus becomes very large as $l' \to l$ and $l' \to 0$.
Semiclasically, fixed a point of the random walk, it is easier for it to coincide with another point of the random walk after a small number of steps (note that both limits imply that one of the daughter strings is short).
The objective of this appendix is to analyze the range of validity of this formula, and thus of the random walk interpretation. \\

In order to obtain Eq.~\eqref{eq:divergent-decay} as the leading contribution to the decay rate, we assumed that the level of the strings involved was much larger than one, and this allowed us to use the Hardy-Ramanujan approximation for the number of states at a given level, $\mathcal{G}(N)$.
The validity of Eq.~\eqref{eq:divergent-decay}, and thus of the random walk interpretation, is dependent on how well the Hardy-Ramanujan approximation performs.
This can be quantified using the calculations of Mañes~\cite{Manes:2001cs}, who computed higher order corrections to the Hardy-Ramanujan approximation.
Naming $\mathcal{G}_c (N,D)$ the corrected expression for the number of oscillators in $D$ spacetime dimensions,~\cite{Manes:2001cs} finds:

\begin{equation}
    \mathcal{G}_c(N,D)\sim \lr{\frac{24}{D-2}}^{1/2}\frac{e^{a(x)}}{(2\pi b(x))^{1/2}}\lr{1+\frac{c(x)}{D-2}}\, , 
\end{equation}
\begin{gather*}
    a(x)\equiv \lr{N-\frac{D-2}{24}}\frac{1}{x}+\frac{D-2}{24}(4\pi^2x-12\log (2\pi x))\, ,\\
    b(x)\equiv 4x^2(2\pi^2x-3)\, , \\
    c(x)\equiv \frac{9}{2}\frac{4\pi^2x-3}{(2\pi^2x-3)^2}-\frac{45(\pi^2x-1)^2}{(2\pi^2x-3)^3}\, , \\
    x\equiv \frac{1}{2\pi^2}\lr{3+\sqrt{9+\pi^2 \lr{\frac{24N}{D-2}-1}}}\, .
\end{gather*}

We can have a taste of the accuracy of this expression by noting that it gives good results even for $N=1$ in $D=10$, giving $\mathcal{G}_c(1,10)=8.029$ against the true result, $8$. 
We will take $\mathcal{G}_c$ as the true value of the number of states, and compare it to the result obtained using the Hardy-Ramanujan expression to check the well-known range of validity of the approximation, $N\geq 5$.
To do so, we have compared the relative error $|\mathcal{G}(N)-\mathcal{G}_c(N,10)|/\mathcal{G}_c(N,10)$ numerically to deduce that the expressions agree within a relative error smaller than $10\%$ for all $N\geq 5$ (in $D=10$).

We therefore conclude that we can set the cutoff at this value, or in terms of the fundamental string length $l_c=4\sqrt{5}l_s\sim 9l_s$.
We interpret this result as implying that the strings are well approximated by random walks at scales larger than around an order of magnitude higher than the string scale.
Analogously, semiclassical fundamental strings develop nontrivial small scale structure at such scales.
This could be interesting in the context of cosmic strings.
\section{Out of equilibrium: details}\label{sec:detail-perturbations}
In this Appendix we study the details of the terms appearing in the linearized Boltzmann equation to argue for the behaviour of the fluctuations, thus inferring equilibration rates.
\subsection*{Solutions in zero dimensions}
We can convert the integral equation in~\eqref{eq:diffeq-f} into a second order PDE as follows.
Consider the term
\begin{equation}
    G(l,t)\equiv \int_0^l{dl'\, l'\delta n(l',t)\lr{e^{\frac{-(l-l')}{L}}-1}}\, .
\end{equation}
Taking its derivative, we find:
\begin{equation}
    \frac{\partial G(l,t)}{\partial l}=\frac{-1}{L}\int_0^l{dl'\, l'\delta n(l',t)e^{\frac{-(l-l')}{L}}}=\frac{-1}{L}\lr{G(l,t)+\int_0^l{dl'\, l'\delta n(l',t)}}\, .
\end{equation}
Where we have used the definition of $G(l,t)$ in the second equality.
The key observation is that there is no $l-$dependence in the integrand, and thus taking a further derivative we find:
\begin{equation}\label{eq:diffeq-G}
    L \frac{\partial ^2 G(l,t)}{\partial l^2}+\frac{\partial G(l,t)}{\partial l}=-l\delta n(l,t)\, .
\end{equation}
Taking derivatives of Eq.~\eqref{eq:diffeq-f} and introducing them into~\eqref{eq:diffeq-G}, we find the following condition:
\begin{equation}
    \left[2(l+L)+\lr{\frac{l^2}{2}+l L+\frac{V}{\kappa}\frac{\partial}{\partial t}}\frac{\partial }{\partial l}\right] \lr{\delta n(l,t)+L \frac{\partial \delta n}{\partial l}}=0 \, .
\end{equation}
And thus the problem can be divided into two simpler problems: first, find the kernel of the operator
\begin{equation}
    \mathcal{L}\equiv 2(l+L)+\lr{\frac{l^2}{2}+l L+\frac{V}{\kappa}\frac{\partial}{\partial t}}\frac{\partial }{\partial l} \, ,
\end{equation}
and then translate the functions in the kernel, denoted $K(l,t)$, into fluctuations through a first order inhomogeneous ODE:
\begin{equation}
    \delta n(l,t)+ L \frac{\partial \delta n (l,t)}{\partial l}= K(l,t) \, ,
\end{equation}
which is solved by:
\begin{equation}
\delta n(l,t)= \frac{e^{-l/L}}{L}\int_0^l{dl'\, e^{l'/L}K(l',t)}+g(t)e^{-l/L}\, ,
\end{equation}
where $g(t)$ is an arbitrary function of time.
Before going ahead and computing more interesting solutions, notice that there is a trivial solution satisfying $K(l,t)=0$.
In terms of the fluctuation, this is simply $\delta n(l,t)=e^{-l/L}\propto l\Bar{n}(l)$ (c.f: Eq.~\eqref{eq:eq-dist-closed}).
More interesting solutions are those with $K(l,t)\neq 0$. 
Let us discuss them.
\\ 

First, notice that not all functions in the kernel of $\mathcal{L}$ are (upon integration) solutions of Eq.~\eqref{eq:diffeq-f}.
An obvious example is $g(t)e^{-l/L}$ with nonconstant $g(t)$.
We have found more complicated functions in the kernel, which do not actually solve the original equation.

Nevertheless, let us discuss a true family of solutions to Eq.~\eqref{eq:diffeq-f}.
One can easily check that the function
\begin{equation}
    K(l,t)=(c /L^2+t)\frac{\kappa}{V}e^{-\frac{\kappa}{V}\lr{\frac{l^2}{2}+lL}(c/L^2+t)}\, 
\end{equation}
is in the kernel of $\mathcal{L}$ for arbitrary real $c$ (and of course an arbitrary multiplicative constant which we have chosen for convenience).
In terms of the perturbation, this reads:
\begin{equation}\label{eq:analytic-fluctuation}
    \delta n(l,t)=\sqrt{\frac{\kappa}{V}}\sqrt{\frac{\pi(c+t L^2)}{2}}\frac{e^{-\frac{l}{L}+A(t)^2}}{L^2}\text{Erf}\lr{A(t),\sqrt{\frac{\kappa }{V}}\sqrt{\frac{c+t L^2}{2}}\frac{l}{L}+A(t)}\, ,
\end{equation}
where $\text{Erf}(z_1,z_2)=\frac{2}{\sqrt{\pi}}\int_{z_1}^{z_2}{e^{-t^2}dt}$ is the incomplete error function, and
\begin{equation}
    A(t)=\sqrt{\frac{V}{\kappa }}\sqrt{\frac{c+t L^2}{2}}\lr{\frac{\kappa }{V}-\frac{1}{c+tL^2}}\, .
\end{equation}

\subsection*{Spreading of fluctuations}
Let us now study the case $d\neq 0$.
Our objective is to show that spreading is supressed for localized fluctuations.
The linearized Boltzmann equation now reads:
\begin{equation}
    \begin{split}
     \frac{\partial \lr{\delta n(l,t)}}{\partial t}= & 
     \kappa\int_{l_c}^{l-l_c}{\frac{dl'}{\lr{l-l'}^{d/2}}\, \lr{e^{-(l-l')/L}l'\delta n(l',t)-\frac{1}{2}\lr{\frac{l}{l'}}^{d/2}l\delta n(l,t) }}
     \\ & 
     +\kappa \int_{l+l_c}^\infty{dl' \, \lr{ l'\delta n(l',t)\lr{\frac{l'}{l(l'-l)}}^{d/2}-l\delta n(l,t)\frac{e^{-(l'-l)/L}}{(l'-l)^{d/2}}}} \, .
\end{split}
\end{equation}

Let us first argue that, as before, the propagation of a fluctuation in length space is suppressed, \textit{i.e}, that a fluctuation that initially has a compact support of spread $\Delta l$ around $l_1$ contributes to $\delta n (l,t),\,  l\notin \Delta l$ at low order in $1/L$.
The strategy is to show that, far from the location of the perturbation, the leading-order contribution to $\partial \delta n(l,t)/\partial t$ is proportional to the energy of the perturbation, $\int_{\Delta l_p}{dl' \, l'\delta n(l',t)}$, which is zero for fluctuations of interest (any fluctuation with nonzero energy is a sum of $l\bar n(l)$ with a zero-energy fluctuation).

For short strings, $l\ll l_1$, the nonzero contribution is given by 
\begin{equation}
\kappa\int_{l+l_c}^{\infty}{dl'\, l'\delta n(l',t)\lr{\frac{l'}{l(l'-l)}}^{d/2}}
\end{equation}
However, since $l'\geq l_1\gg l$, the term in brackets is approximately constant, giving a contribution
\begin{equation}
    \frac{\partial \lr{\delta n(l,t)}}{\partial t}= 
    \frac{\kappa}{l^{d/2}} \int_{\Delta l_p}{dl' \,  l'\delta n(l',t)\lr{1+\frac{d\, l}{2l'}+\mathcal{O}\lr{\frac{l}{l'}}^2}}\simeq\frac{d\, \kappa}{2\, l^{d/2-1}}\delta N(t) \, ,
\end{equation}
where $\delta N(t)\equiv \int_{\Delta l}{dl\, \delta n(l,t)}$ is the total number of strings involved in the fluctuation.
Thus, such localized perturbations do not affect the shorter strings at first order in $(l/l_1)$, and we keep the higher-order term because of its interesting $d$-dependence, which might or might not prefer to populate very short strings.
\\

Let us now move to long strings. 
The only relevant term is now 
\begin{equation}
\kappa\int_{l_c}^{l-l_c}{dl'\, e^{-(l-l')/L}\frac{l'\delta n(l',t)}{(l-l')^{d/2}}}
\end{equation}
In this case, we can expand for small $l'/l$ to find:
\begin{equation}
\frac{\partial \delta n(l,t)}{\partial t} = \kappa \frac{e^{-l/L}}{l^{d/2}}\int_{\Delta l_p}{dl' \, l'e^{l'/L}\delta n(l',t)\lr{1+\mathcal{O}\lr{\frac{l'}{l}}}}\, .
\end{equation}
Thus, at first order in $l'/l$ we find a Boltzmann suppression factor times the energy of the perturbation, if $\Delta l\ll L$ so we can take the exponential
in the integrand to be approximately constant. 
In the case of fluctuations, this is zero, and the leading contribution to spreading is supressed.
We have thus shown that spreading of a localized fluctuation is small (although it cannot be zero since the energy of the fluctuation needs to be preserved).
We use this argument in the main text to estimate equilibration rates from the decaying terms.

\bibliographystyle{utphys}
\bibliography{biblio}
\end{document}